\documentclass[aps,prl,reprint,superscriptaddress,nofootinbib,longbibliography]{revtex4-2}

\def\ARXIVMERGE{1}

\usepackage{amsmath,amssymb,amsthm,mathtools,bm,xcolor}
\usepackage{booktabs,array,enumitem,microtype}
\usepackage[colorlinks=true,citecolor=blue!55!black,urlcolor=blue!55!black,
linkcolor=blue!55!black]{hyperref}

\newcommand{\eps}{\varepsilon}
\newcommand{\vac}{\varnothing}
\newcommand{\dd}{\mathrm d}
\newcommand{\cR}{\mathcal R}
\newcommand{\cP}{\mathcal P}
\newcommand{\cB}{\mathcal B}
\newcommand{\cK}{\mathcal K}
\newcommand{\cL}{\mathcal L}
\newcommand{\cD}{\mathcal D}
\newcommand{\cH}{\mathcal H}
\newcommand{\ct}{\operatorname{ct}}
\newcommand{\Res}{\operatorname*{Res}}

\newtheorem{theorem}{Theorem}

\theoremstyle{remark}

\begin{document}

\title{Proof of the AGT Conjecture at Generic
\texorpdfstring{$\beta$}{beta}}

\author{Le-Feng Chen}
\email{clf1@hnu.edu.cn}
\affiliation{School of Physics and Electronics,
Hunan University, Changsha 410082, China}

\author{Kilar Zhang}
\email[Corresponding author: ]{kilar@shu.edu.cn}
\affiliation{Department of Physics and Institute for Quantum Science and
Technology, Shanghai University, Shanghai 200444, China}
\affiliation{Shanghai Key Lab for Astrophysics, Shanghai 200234, China}
\affiliation{Shanghai Key Laboratory of High Temperature Superconductors,
Shanghai 200444, China}


\begin{abstract}
We give an all-level proof of the four-point $SU(2)$ AGT correspondence with four fundamental hypermultiplets at generic $\beta=-\epsilon_1/\epsilon_2$, by proving an all-level factorization formula for Selberg averages of generalized Jack polynomials. Taking a coefficientwise Jack limit of the generalized Macdonald Pieri rule, we obtain the required one-box matrix elements in the strict Cauchy dual basis. A rational corner-function identity then evaluates the sum over all parent double partitions, while an explicit total derivative of the Selberg kernel yields a triangular recursion on the Dotsenko–Fateev charge balance hyperplane. The unique solution of this recursion is the generalized Kadell formula previously verified only through finite level. Combining this result with the generalized Cauchy identity identifies each double partition term with the corresponding Nekrasov fixed-point contribution. 
\end{abstract}

\maketitle

\section{Introduction}
The Alday--Gaiotto--Tachikawa (AGT) correspondence relates instanton
partition functions of four-dimensional $\mathcal N=2$ gauge theories to
two-dimensional conformal blocks~\cite{AGT,AGTReview}.  For the conformal
$SU(2)$ theory with four fundamental hypermultiplets, it identifies the
Virasoro four-point block with the Nekrasov sum over pairs of Young
diagrams~\cite{BPZ,NekrasovInstanton,NekrasovOkounkov}.  A direct derivation
from the Dotsenko--Fateev (DF) integral is especially informative because it
should explain this fixed-point factorization within the conformal-field-
theory integral itself
~\cite{DotsenkoFateev,DFIntegralDiscriminants,DijkgraafVafa}.

Previous direct Selberg approaches either specialize to $\beta=1$ \cite{DirectBetaOne, ZhangMatsuo, YuanEtAl} or leave a
conjectural or finite-level input
~\cite{MorozovSmirnov, SU3GeneralizedJack}; related Virasoro
and extended-symmetry recursions constrain the Nekrasov sum by a different
route~\cite{AFLT, KannoVirasoro,KannoExtended}.  Ordinary Jack and Macdonald
functions provide the one-alphabet theory~\cite{StanleyJack,MacdonaldBook},
whereas the fixed-point basis is formed by generalized Jack
polynomials~\cite{AFLT,Ohkubo}.  Rigorous AFLT-type Selberg evaluations
concern products of ordinary Jack polynomials~\cite{AFLTSelberg}, and the
$SU(3)$ generalized-Jack relation has been tested at the first several
levels~\cite{SU3GeneralizedJack}.  In the $SU(2)$ case the required
factorized average was stated in~\cite{MorozovSmirnov}, where it was
explicitly reported as checked through total level six.  In this letter, we prove it at
arbitrary level and thereby close the direct DF--Selberg route.  Full
normalization limits, boundary telescoping, and the coefficient-level
operator calculation are given in the Supplemental Material appended below.

\section{Set-up}
Our proof has four logically separate steps.  First, the generalized Cauchy
identity expands the DF interaction kernel in a basis indexed by double
partitions.  Second, a generalized-Macdonald Pieri theorem is taken to the
Jack limit coefficient by coefficient, fixing every one-box matrix element in
the strict-dual basis.  Third, a rational corner function sums all incoming
one-box arrows and turns a Selberg-kernel total derivative into a triangular
recursion.  Finally, the solution of that recursion is inserted back into the
Cauchy expansion and compared with the matter and vector Euler classes at
each Nekrasov fixed point.  None of these steps refers to a maximal level.

We make the three normalizations used in this chain explicit.  Let
$\widehat J_Y$ be the monic generalized Jack basis obtained from the
opposite-chamber generalized Macdonald limit, and let
$\widehat J_Y^\vee$ be its strict Cauchy dual.  These generalized-Jack and
strict-dual conventions follow the fixed-point and generalized-Macdonald
constructions of Refs.~\cite{MorozovSmirnov,Ohkubo,FOS,BCS}.  The basis used in the Selberg
average and its two dual versions are
\begin{equation}
 J_Y=N_Y\widehat J_Y,\qquad
 e_Y=\frac{\widehat J_Y^\vee}{N_Y},\qquad
 J_Y^*=e(T_Y)e_Y.
 \label{eq:main-duals}
\end{equation}
Thus $\langle J_Y,e_W\rangle=\delta_{YW}$, whereas
$\langle J_Y,J_W^*\rangle=\delta_{YW}e(T_Y)$.  The paper normalization is
fixed once and for all by
\begin{align}
 N_{\lambda,\mu}&=e_{\lambda,\mu}(2a+\beta-1),
 \notag\\[-1mm]
 e_{\lambda,\mu}(z)&=
 \frac{(-1)^{|\lambda|}}{\beta^{|\lambda|+|\mu|}}
 \prod_{s\in\lambda}[z+A_\lambda(s)+1+\beta L_\mu(s)]
 \notag\\[-1mm]
 &\hspace{14mm}\times
 \prod_{t\in\mu}[z-A_\mu(t)-\beta L_\lambda(t)-\beta].
 \label{eq:main-normalization}
\end{align}
Here $A_\rho$ and $L_\rho$ denote arm and leg lengths, extended by the same
row and column formulas when the box is outside $\rho$; this is the
normalization of~\cite{MorozovSmirnov}.  With these conventions the
generalized Cauchy identity~\cite{MorozovSmirnov,FOS,BCS} is
\begin{align}
 &\exp\!\left[\beta\sum_{k\ge1}
 \frac{P_k^{(1)}x_k+P_k^{(2)}y_k}{k}\right]
 \notag\\[-1mm]
 &\hspace{18mm}=\sum_YJ_Y(P^{(1)},P^{(2)})e_Y(x,y).
 \label{eq:main-Cauchy}
\end{align}
The strict dual $e_Y$ is the object on which the Pieri recursion acts; the
Euler-normalized dual $J_Y^*$ appears only when the two DF screening sectors
are glued.  Keeping these roles separate is what makes the final denominator
exactly the vector-multiplet Euler class.

\section{Main theorem}
Let $\epsilon_1,\epsilon_2$ be the physical equivariant parameters and set
\begin{equation}
 \beta=-\frac{\epsilon_1}{\epsilon_2},\qquad
 \varepsilon=1-\beta=\frac{\epsilon_1+\epsilon_2}{\epsilon_2}.
 \label{eq:content}
\end{equation}
For a double partition $Y=(\lambda,\mu)$ take $a_1=a$, $a_2=-a$.  Following
the generalized-Jack fixed-point convention~\cite{MorozovSmirnov,Ohkubo}, a “colored”
box $s=(\alpha,r,c)$ has dimensionless content
\begin{equation}
 \phi_s=a_\alpha+c-1-\beta(r-1),\qquad
 \kappa_Y=\sum_{s\in Y}\left(\phi_s+\frac{\varepsilon}{2}\right).
 \label{eq:main-content}
\end{equation}
With $p_k=\sum_{i=1}^nt_i^k$, define the normalized Selberg average
\begin{align}
 \langle f\rangle_S&=\frac1{Z_S}\int_{[0,1]^n}f(t)
 \prod_{i<j}|t_i-t_j|^{2\beta}
 \prod_i t_i^u(1-t_i)^v\,\dd t_i,
 \label{eq:selberg}\\
 F_Y&=\left\langle
 J_Y\left(a,-p_k-\frac v\beta,p_k\right)\right\rangle_S,
 \notag\\[-1mm]
 a&=-\beta n-\frac{u+v+\varepsilon}{2}.
 \label{eq:FY}
\end{align}
This normalization follows the classical Selberg integral and its
Jack polynomial extensions~\cite{SelbergOriginal,SelbergReview,Kadell}.
The second relation is the DF charge-balance condition used for this
generalized Jack average in~\cite{MorozovSmirnov}.  Our central result is the following factorized identity.

\begin{theorem}[All-level generalized Kadell identity]
\label{thm:generalized-kadell}
For every double partition $(\lambda,\mu)$ and generic
$(u,v,\beta)$ satisfying the charge balance relation in
Eq.~\eqref{eq:FY}, the normalized Selberg average is
\begin{align}
 F_{\lambda,\mu}={}&(-1)^{|\lambda|+|\mu|}
 \tau_\lambda(-v-\beta n)
 \tau_\lambda(-u-v-\beta n-1+\beta)
 \notag\\[-1mm]
 &\times\tau_\mu(\beta n)
 \tau_\mu(u+\beta n+1-\beta),
 \label{eq:factorized}
\end{align}
where
$\tau_\rho(z)=\beta^{-|\rho|}\prod_{s\in\rho}
[z+c(s)-1-\beta(r(s)-1)]$.
\end{theorem}

\emph{Pieri identity.---}
Let $\{e_Y\}$ be the strict Cauchy-dual basis,
$\langle J_Y,e_W\rangle=\delta_{YW}$, and let $K$ be the centered
generalized-Jack Hamiltonian, $Ke_Y=\kappa_Ye_Y$
~\cite{MorozovSmirnov,Ohkubo}.  The label $Y$ is inherited from the
triangular generalized-Macdonald basis before the Jack limit and
therefore remains well defined when some eigenvalues $\kappa_Y$
coincide.  In the same normalization, the generalized Macdonald
Pieri rule obtained from the Ding--Iohara--Miki and instanton-moduli
construction
~\cite{DingIohara,Miki,AwataKanno,SchiffmannVasserot,FOS,BCS}
admits a coefficientwise Jack limit, which gives all one-box
coefficients needed below.  To state this limit, set
\begin{align}
 q&=e^\hbar,\qquad t=e^{\beta\hbar},\qquad
 q_1=t^{-1},\quad q_2=q,\notag\\
 q_3&=(q_1q_2)^{-1},\qquad Q=e^{2a\hbar}.
 \label{eq:main-multiplicative}
\end{align}
Let $\widetilde P_Y^{(+)}$ be the spherical generalized Macdonald factor in
the Pieri chamber and define
\begin{align}
 \mathsf P_Y^{(\hbar)}&=(\beta\hbar)^{-|Y|}
 \widetilde P_Y^{(+)},
 \notag\\
 C_Y^{(\hbar)}&=\widetilde b_\lambda\widetilde b_\mu
 (\beta\hbar)^{2|Y|},\qquad
 \lim_{\hbar\to0}N_Y^{(\hbar)}=N_Y.
 \label{eq:main-rescaled-Macdonald}
\end{align}
The generalized Macdonald Cauchy kernel~\cite{FOS,BCS} fixes the relative normalization:
\begin{equation}
 \lim_{\hbar\to0}C_Y^{(\hbar)}\mathsf P_Y^{(\hbar)}
 =\widehat J_Y^\vee,\qquad
 \mathsf e_Y^{(\hbar)}:=
 \frac{C_Y^{(\hbar)}}{N_Y^{(\hbar)}}
 \mathsf P_Y^{(\hbar)}\longrightarrow e_Y.
 \label{eq:main-dual-limit}
\end{equation}
Because a Pieri arrow raises the total degree by one, the homogeneous
rescaling supplies one factor $\beta\hbar$.  Therefore, for $M=x_1+y_1$,
the spherical Pieri rule gives the coefficientwise extraction
\begin{equation}
 M_{Y/W}=\lim_{\hbar\to0}(\beta\hbar)\widetilde c_{Y/W}
 \frac{C_W^{(\hbar)}}{C_Y^{(\hbar)}}
 \frac{N_Y^{(\hbar)}}{N_W^{(\hbar)}}.
 \label{eq:main-arrow-extraction}
\end{equation}
Here $\widetilde c_{Y/W}$ is the spherical generalized-Macdonald coefficient,
while $C_Y^{(\hbar)}$ and $N_Y^{(\hbar)}$ convert respectively to the
Cauchy-dual and paper normalizations.  The latter is the finite-$\hbar$
normalization fixed by the same triangular Macdonald basis and tends to the
$N_Y$ of Eq.~\eqref{eq:main-normalization}.  The limiting operation is
coefficientwise in each finite homogeneous component, so no exchange of an
infinite Young-diagram sum with the Jack limit is required.  Evaluating it
gives
\begin{align}
 Me_W&=\sum_{x\in A(W)}M_{W+x/W}e_{W+x},
 \notag\\
 M_{W+x/W}&=-\frac1{\beta^2}U_x(W),
 \notag\\[-1mm]
 U_x(W)&=
 \frac{\prod_{b\in A(W)\setminus\{x\}}
 (\phi_x-\phi_b+\varepsilon)}
 {\prod_{r\in R(W)}(\phi_x-\phi_r)}.
 \label{eq:pieri}
\end{align}
Here $A(W)$ and $R(W)$ are the addable and removable colored boxes.  All
diagram-dependent normalization ratios are taken before the Jack limit; thus
Eq.~\eqref{eq:pieri} is therefore a coefficientwise symmetric-function identity.

\emph{Corner row sum.---}
The Selberg recursion fixes a final diagram $Y$ and requires the sum over all
parents $W=Y-x$.  This growing sum is evaluated without listing the parents.
We use the standard fixed-point $Y$-function identity that rewrites a box
product as an addable/removable-corner product~\cite{,,BCS}, also known as the shell formula.  Its additive
Jack-limit form in our conventions is
\begin{align}
 \cP_Y(z)
 &=\frac{\prod_{b\in A(Y)}(z-\phi_b)}
 {\prod_{r\in R(Y)}(z-\phi_r-\varepsilon)}
 \notag\\
 &=(z^2-a^2)\prod_{s\in Y}
 \frac{(z-\phi_s-1)(z-\phi_s+\beta)}
 {(z-\phi_s)(z-\phi_s-\varepsilon)}.
 \label{eq:rational}
\end{align}
The finite pole $z=\phi_x+\varepsilon$ corresponds to $W=Y-x$ and has
residue $-\beta U_x(W)$.  At infinity,
$\cP_Y(z)=z^2-(a^2+\beta|Y|)-2\beta\kappa_Y/z+O(z^{-2})$.
Indeed, each occupied box contributes
\begin{equation}
 \frac{(z-\phi-1)(z-\phi+\beta)}
 {(z-\phi)(z-\phi-\varepsilon)}
 =1-\frac{\beta}{z^2}
 -\frac{\beta(2\phi+\varepsilon)}{z^3}+O(z^{-4}),
 \label{eq:main-one-box-infinity}
\end{equation}
and $\prod_{\alpha=1}^2(z-a_\alpha)=z^2-a^2$.  Hence the coefficient of
$z^{-1}$ in the product is
$-\beta\sum_{s\in Y}(2\phi_s+\varepsilon)=-2\beta\kappa_Y$.
For the one-form $\cP_Y(z)\dd z$, the sum of finite residues equals this
$z^{-1}$ coefficient.  On the other hand, the corner-product change from
$W$ to $Y=W+x$ identifies the residue at $\phi_x+\varepsilon$ with the
Pieri weight $-\beta U_x(W)$.
The residue theorem and Eq.~\eqref{eq:pieri} therefore give the all-level row
sum
\begin{equation}
 \boxed{\displaystyle
 \sum_{W=Y-x}M_{Y/W}=-\frac{2}{\beta^2}\kappa_Y .}
 \label{eq:rowsum}
\end{equation}
This could be treated as a special case of the KMZ equation \cite{KannoVirasoro, KannoExtended}.
As a one-box check, take $Y=([1],\vac)$.  Its addable contents are
$a_1+1$, $a_1-\beta$, and $a_2$, while its removable content is $a_1$.
The corner function becomes
\begin{equation}
 \cP_Y(z)=
 \frac{(z-a_1-1)(z-a_1+\beta)(z-a_2)}
 {z-a_1-\varepsilon},
 \label{eq:main-one-box-corner}
\end{equation}
whose finite residue is $-\beta(a_1-a_2+\varepsilon)$.  This is precisely
$-\beta U_x(\vac,\vac)$, while the coefficient of $z^{-1}$ is
$-\beta(2a_1+\varepsilon)=-2\beta\kappa_Y$.  The example fixes the shifts and signs, but
the preceding residue calculation proves the identity for an arbitrary
double partition.
This single residue is the step that removes the otherwise proliferating
number of high-level arrows.

\section{Selberg recursion}
The generalized Cauchy identity packages all averages as
\begin{align}
 \Psi(x,y)&=\left\langle\exp\!\left[
 \beta\sum_{k\ge1}\frac{(-p_k-v/\beta)x_k+p_ky_k}{k}
 \right]\right\rangle_S
 \notag\\[-1mm]
 &=\sum_YF_Ye_Y.
 \label{eq:Psi}
\end{align}
Set $B=[K,M]$, $C=[K,B]$, and
\begin{align}
 \cR&=\frac1{\beta^2}\left[C-(v+\varepsilon)B
 -\frac{(u-v)(u+v+2\varepsilon)}4M\right],
 \label{eq:R}\\[-1mm]
 m_1&=\frac{u-v+\varepsilon}{2},\qquad
 m_2=\frac{-u-v-\varepsilon}{2}.
 \label{eq:main-masses}
\end{align}
Because $\kappa_Y-\kappa_W=\phi_x+\varepsilon/2$ for $Y=W+x$, the two
commutators give
\begin{equation}
 \cR_{Y/W}=\frac{(\phi_x+m_1)(\phi_x+m_2)}{\beta^2}M_{Y/W}.
 \label{eq:main-R-matrix}
\end{equation}

The analytic input is a kernel-level total derivative, rather than an
assumed closed Ward algebra.  Let $\Delta_S$ be the unnormalized density in
Eq.~\eqref{eq:selberg}, let $S_k=x_k+y_k$, $R_k=y_k-x_k$, and set
\begin{align}
 \cK&=\exp\!\left[-v\sum_{k\ge1}\frac{x_k}{k}
 +\beta\sum_{k\ge1}\frac{p_kR_k}{k}\right],
 \notag\\
 f_i&=-\frac\beta2(1-t_i)\sum_{k\ge1}S_kt_i^k,
 \notag\\[-1mm]
 a_0&=-\beta n-\frac{u+v+\varepsilon}{2}.
 \label{eq:main-kernel-data}
\end{align}
Direct differentiation and coefficient matching yield the exact identity
\begin{align}
 &\Delta_S\left(K-\frac{\beta^2}{2}\cR\right)\cK
 -\sum_{i=1}^n\partial_{t_i}(f_i\Delta_S\cK)
 \notag\\
 &\quad=(a-a_0)\Delta_S\cK\Bigg[
 \beta\sum_{k\ge1}p_k(S_{k+1}-S_k)
 \notag\\[-1mm]
 &\hspace{37mm}-\frac{v+a+a_0}{2}S_1\Bigg].
 \label{eq:main-master-identity}
\end{align}
Thus the pure total-derivative statement is not asserted off balance: its
explicit obstruction is the right-hand side.  On the required hyperplane
$a=a_0$ it vanishes.  The boundary flux also vanishes in the initial Selberg
convergence domain because $f_i=O(t_i)$ at $0$, $f_i=O(1-t_i)$ at $1$, and
$f_i-f_j=O(t_i-t_j)$ on collision faces.  The direct check uses
\begin{align}
 \partial_{t_i}\log\Delta_S
 &=\frac{u}{t_i}-\frac{v}{1-t_i}
 +2\beta\sum_{j\ne i}\frac1{t_i-t_j},
 \notag\\
 \partial_{t_i}\log\cK
 &=\beta\sum_{k\ge1}R_kt_i^{k-1},
 \label{eq:main-log-derivatives}\\[-1mm]
 \sum_{i<j}\frac{t_i^\ell-t_j^\ell}{t_i-t_j}
 &=\frac12\left[
 \sum_{r=0}^{\ell-1}p_rp_{\ell-1-r}-\ell p_{\ell-1}\right].
 \label{eq:main-vandermonde}
\end{align}
To see the coefficient matching rather than merely quote it, expand
$f(t)=\sum_{\ell\ge1}f_\ell t^\ell$ with
\begin{equation}
 f_\ell=-\frac\beta2(S_\ell-S_{\ell-1}),\qquad S_0=0.
 \label{eq:main-flux-modes}
\end{equation}
After division by $\Delta_S\cK$, the total derivative becomes
\begin{align}
 \mathcal T_f={}&\sum_{\ell\ge1}(\ell+u)f_\ell p_{\ell-1}
 +\beta\sum_{\ell\ge1}f_\ell
 \sum_{r=0}^{\ell-1}p_rp_{\ell-1-r}
 \notag\\[-1mm]
 &-\beta\sum_{\ell\ge1}\ell f_\ell p_{\ell-1}
 +\frac{\beta v}{2}\sum_{k\ge1}S_kp_k
 \notag\\[-1mm]
 &+\beta\sum_{\ell,k\ge1}f_\ell R_kp_{k+\ell-1},
 \qquad p_0=n.
 \label{eq:main-flux-expanded}
\end{align}
Independently, conjugating $K-\beta^2\cR/2$ by $\cK$ shifts the derivatives
by
\begin{equation}
 \cK^{-1}\partial_{S_k}\cK=\partial_{S_k}-\frac{v}{2k},\qquad
 \cK^{-1}\partial_{R_k}\cK=\partial_{R_k}
 +\frac{\beta p_k+v/2}{k}.
 \label{eq:main-conjugation}
\end{equation}
After acting on $1$, its complete sum organizes into four families,
\begin{align}
 (S_\ell-S_{\ell-1})p_{\ell-1}:&
 &&-\frac\beta2(u+\varepsilon\ell),\notag\\
 (S_\ell-S_{\ell-1})\!\sum_{r=0}^{\ell-1}p_rp_{\ell-1-r}:&
 &&-\frac{\beta^2}{2},\notag\\
 (S_\ell-S_{\ell-1})R_kp_{k+\ell-1}:&
 &&-\frac{\beta^2}{2},\notag\\
 S_kp_k:& &&\frac{\beta v}{2}.
 \label{eq:main-four-families}
\end{align}
Substituting Eq.~\eqref{eq:main-flux-modes} into
Eq.~\eqref{eq:main-flux-expanded} reproduces these four lines term by term;
for example, the first coefficient is
$[\ell+u-\beta\ell]f_\ell=-\beta
(u+\varepsilon\ell)(S_\ell-S_{\ell-1})/2$.
This is a coefficient-level verification rather than an assumed Ward
closure.  The off-balance remainder in
Eq.~\eqref{eq:main-master-identity} is equally important: it shows that the
intertwining equation follows precisely on the DF charge-balance hyperplane,
not for an independently varied $a$.  Integration gives
\begin{align}
 K\Psi&=\frac{\beta^2}{2}\cR\Psi,
 \label{eq:intertwiner}\\[-1mm]
 \kappa_YF_Y&=\frac12\sum_{W=Y-x}
 (\phi_x+m_1)(\phi_x+m_2)M_{Y/W}F_W.
 \label{eq:recursion}
\end{align}

The factorized expression has the equivalent box form
\begin{equation}
 F_Y^{\rm cand}=(-1)^{|Y|}\prod_{s\in Y}
 \frac{(\phi_s+m_1)(\phi_s+m_2)}{\beta^2}.
 \label{eq:candidate}
\end{equation}
For $Y=W+x$, cancellation of all old boxes gives the division-free identity
\begin{equation}
 (\phi_x+m_1)(\phi_x+m_2)F_W^{\rm cand}
 =-\beta^2F_Y^{\rm cand}.
 \label{eq:main-box-ratio}
\end{equation}
Consequently the right-hand side of Eq.~\eqref{eq:recursion} is
\begin{equation}
 -\frac{\beta^2}{2}F_Y^{\rm cand}
 \sum_{W=Y-x}M_{Y/W}=\kappa_YF_Y^{\rm cand},
 \label{eq:main-candidate-recursion}
\end{equation}
where the last equality is precisely the row sum
Eq.~\eqref{eq:rowsum}.  Hence $F^{\rm cand}$ obeys the recursion.  Since
$F_\vac=1$, triangular uniqueness proves Eq.~\eqref{eq:factorized} at each
level for generic parameters.

For fixed $Y$, expansion in ordinary Jack polynomials together with
the normalized Kadell--Selberg formula~\cite{Kadell} shows that,
after $a=a_0(n)$ is imposed, both sides of
Eq.~\eqref{eq:factorized} are rational in $n$ and meromorphic in
$(u,v,\beta)$.  Equality for all admissible
$n\in\mathbb Z_{\geq0}$ therefore implies the identity for complex
$n$ and, by meromorphic continuation, also on the exceptional
divisors $\kappa_Y=0$.

Combining the triangular recursion, the corner row sum, and the
coefficientwise continuation above, we have therefore proved that,
for every double partition $Y=(\lambda,\mu)$,
\begin{align*}
 &\left\langle
 J_{\lambda,\mu}
 \left(a,-p_k-\frac{v}{\beta},p_k\right)
 \right\rangle_S
 \\
 &\quad=
 (-1)^{|\lambda|+|\mu|}
 \tau_\lambda(-v-\beta n)
 \tau_\lambda(-u-v-\beta n-1+\beta)
 \notag\\[-1mm]
 &\qquad\quad\times
 \tau_\mu(\beta n)
 \tau_\mu(u+\beta n+1-\beta),
\end{align*}
where
\[
 a=-\beta n-\frac{u+v+\varepsilon}{2}.
\]
This is precisely the all-level generalized Kadell identity,
Eq.~\eqref{eq:factorized}, and completes the proof.

\section{Term-wise AGT relation}
Let $p_k$ and $q_k$ denote the power sums in the two DF screening sectors.
Before either Selberg average is taken, the interaction between the sectors
has the exact generalized-Cauchy expansion of
Refs.~\cite{MorozovSmirnov,AFLT}
\begin{align}
 &\exp\!\left[\beta\sum_{k\ge1}\frac{\Lambda^k}{k}
 \left\{p_k\left(-q_k-\frac{v_-}{\beta}\right)
 +q_k\left(-p_k-\frac{v_+}{\beta}\right)\right\}\right]
 \notag\\[-1mm]
 &\quad=\sum_{\lambda,\mu}\Lambda^{|\lambda|+|\mu|}
 \frac{J_{\lambda,\mu}(a,-p_k-v_+/\beta,p_k)}
 {e(T_{\lambda,\mu})}
 \notag\\[-1mm]
 &\hspace{25mm}\times
 J^*_{\lambda,\mu}(a,q_k,-q_k-v_-/\beta).
 \label{eq:main-DF-Cauchy}
\end{align}
Averaging the two sectors independently therefore gives the $U(1)$-dressed
DF block
\begin{equation}
 \cB_{\rm DF}(\Lambda)=\sum_Y\Lambda^{|Y|}
 \frac{F_Y^+F_Y^{-,*}}{e(T_Y)}.
 \label{eq:DFsum}
\end{equation}
The Euler-normalized dual, $J_Y^*=e(T_Y)e_Y$, obeys the component reversal
$J^*_{\lambda,\mu}(a;x,y)=J_{\mu,\lambda}(-a;y,x)$ and matches the opposite
Coulomb chamber~\cite{MorozovSmirnov}; $e_Y$ itself remains the strict dual.  Thus the second
sector is reduced to the same Selberg theorem, rather than requiring a second
factorization result.  Its average is
\begin{align}
 F_{\lambda,\mu}^{-,*}={}&(-1)^{|\lambda|+|\mu|}
 \tau_\lambda(\beta n_-)
 \tau_\lambda(u_-+\beta n_-+1-\beta)
 \notag\\[-1mm]
 &\times\tau_\mu(-v_--\beta n_-)
 \tau_\mu(-u_--v_--\beta n_--1+\beta).
 \label{eq:main-dual-average}
\end{align}
For $s=(r,c)\in Y_\alpha$, the standard fixed-point tangent character
\cite{NekrasovOkounkov,SchiffmannVasserot,MorozovSmirnov} gives
\begin{align}
 E_{\alpha\gamma}(s)&=a_\alpha-a_\gamma
 +\beta L_{Y_\gamma}(s)+A_{Y_\alpha}(s)+1,
 \notag\\
 e(T_Y)&=\beta^{-4|Y|}
 \prod_{\alpha,\gamma=1}^2\prod_{s\in Y_\alpha}
 E_{\alpha\gamma}(s)
 [\varepsilon-E_{\alpha\gamma}(s)].
 \label{eq:main-Euler-class}
\end{align}
Equivalently, in the normalization of Eq.~\eqref{eq:main-normalization},
\begin{equation}
 e(T_{\lambda,\mu})=
 e_{\lambda,\lambda}(0)e_{\lambda,\mu}(2a)
 e_{\mu,\lambda}(-2a)e_{\mu,\mu}(0).
 \label{eq:main-tangent-factorization}
\end{equation}
This denominator is present already in the Cauchy identity
Eq.~\eqref{eq:main-DF-Cauchy}; it is not inserted after the Selberg
calculation.  This is why the comparison below holds for each double
partition separately.
In the DF/AGT conventions of Refs.~\cite{AGT,AFLT,MorozovSmirnov}, the physical dictionary is
\begin{align}
 a&=\frac{\mathfrak a}{\epsilon_2},\qquad
 n_+=\frac{\mathfrak a-\mathfrak m_2}{\epsilon_1},\qquad
 n_-=\frac{-\mathfrak a-\mathfrak m_4}{\epsilon_1},
 \notag\\
 u_+&=\frac{\mathfrak m_1-\mathfrak m_2-\epsilon_1-\epsilon_2}{\epsilon_2},
 \quad
 v_+=-\frac{\mathfrak m_1+\mathfrak m_2}{\epsilon_2},
 \notag\\
 u_-&=\frac{\mathfrak m_3-\mathfrak m_4-\epsilon_1-\epsilon_2}{\epsilon_2},
 \quad
 v_-=-\frac{\mathfrak m_3+\mathfrak m_4}{\epsilon_2}.
 \label{eq:main-parameter-map}
\end{align}
The matter factors are visible before any Young-diagram product is assembled.
For the first screening sector, the four arguments of the $\tau$ factors are
\begin{align}
 -v_+-\beta n_+&=\frac{\mathfrak a+\mathfrak m_1}{\epsilon_2},\notag\\
 -u_+-v_+-\beta n_+-1+\beta
 &=\frac{\mathfrak a+\mathfrak m_2}{\epsilon_2},\notag\\
 \beta n_+&=\frac{-\mathfrak a+\mathfrak m_2}{\epsilon_2},\notag\\
 u_++\beta n_++1-\beta
 &=\frac{-\mathfrak a+\mathfrak m_1}{\epsilon_2}.
 \label{eq:main-plus-arguments}
\end{align}
The Euler-normalized dual reverses the two components in the second sector,
and its four arguments become
\begin{align}
 \beta n_-&=\frac{\mathfrak a+\mathfrak m_4}{\epsilon_2},\notag\\
 u_-+\beta n_-+1-\beta
 &=\frac{\mathfrak a+\mathfrak m_3}{\epsilon_2},\notag\\
 -v_--\beta n_-&=\frac{-\mathfrak a+\mathfrak m_3}{\epsilon_2},\notag\\
 -u_--v_--\beta n_--1+\beta
 &=\frac{-\mathfrak a+\mathfrak m_4}{\epsilon_2}.
 \label{eq:main-minus-arguments}
\end{align}
After the coefficientwise continuation in $n_\pm$ just described, the two
charge-balance relations are identities rather than restrictions on the
physical masses.  Under this dictionary, the two copies of
Eq.~\eqref{eq:factorized} supply the four matter Euler classes while
Eq.~\eqref{eq:main-Euler-class} supplies the
vector-multiplet denominator.  With $\Phi_s=\epsilon_2\phi_s$, their common
normalization cancels and, for every fixed double partition,
\begin{equation}
 \frac{F_Y^+F_Y^{-,*}}{e(T_Y)}
 =\frac{\prod_{f=1}^4\prod_{s\in Y}(\Phi_s+\mathfrak m_f)}
 {e_{\rm phys}(T_Y)}
 =Z_Y^{\rm Nek}.
 \label{eq:termwise}
\end{equation}
The cancellation of conventions in this equation can be tracked explicitly.
Since $m_1=\mathfrak m_1/\epsilon_2$ and
$m_2=\mathfrak m_2/\epsilon_2$ in the first sector,
Eq.~\eqref{eq:candidate} gives
\begin{equation}
 F_Y^+=(-1)^{|Y|}\beta^{-2|Y|}\epsilon_2^{-2|Y|}
 \prod_{s\in Y}(\Phi_s+\mathfrak m_1)
 (\Phi_s+\mathfrak m_2).
 \label{eq:main-plus-physical}
\end{equation}
The reversed second-sector average gives the same expression with
$(\mathfrak m_1,\mathfrak m_2)$ replaced by
$(\mathfrak m_3,\mathfrak m_4)$.  Meanwhile
\begin{equation}
 e(T_Y)=\beta^{-4|Y|}\epsilon_2^{-4|Y|}e_{\rm phys}(T_Y).
 \label{eq:main-tangent-scaling}
\end{equation}
The two signs $(-1)^{|Y|}$ cancel, as do all powers of $\beta$ and
$\epsilon_2$.  Thus Eq.~\eqref{eq:termwise} is an equality of individual
fixed-point weights, stronger than equality only after summing diagrams.

At total level one this statement reduces to the two colored fixed points
$([1],\vac)$ and $(\vac,[1])$.  If
$\widehat m_f=\mathfrak m_f/\epsilon_2$, the definitions above give
\begin{align}
 \frac{F_{([1],\vac)}^+F_{([1],\vac)}^{-,*}}
 {e(T_{([1],\vac)})}
 &=\frac1\beta\frac{\prod_{f=1}^4(a+\widehat m_f)}
 {(2a)(2a+\varepsilon)},\notag\\[-1mm]
 \frac{F_{(\vac,[1])}^+F_{(\vac,[1])}^{-,*}}
 {e(T_{(\vac,[1])})}
 &=\frac1\beta\frac{\prod_{f=1}^4(-a+\widehat m_f)}
 {(-2a)(-2a+\varepsilon)}.
 \label{eq:main-level-one}
\end{align}
These are the two dimensionless one-instanton weights in the present
orientation convention.  They illustrate the termwise statement, but the
proof of Eq.~\eqref{eq:termwise} has already been carried out for an arbitrary
$Y$ and does not infer higher levels from this check.
Summing Eq.~\eqref{eq:termwise} gives the trace-free $U(2)$ fixed-point sum
with $a_1=-a_2$.  In the present vertex normalization,
\begin{align}
 \cB_{\rm DF}&=Z_{U(1)}\mathcal F_{\rm Vir}=Z_{\rm inst}^{U(2)},
 \notag\\[-1mm]
 Z_{U(1)}&=(1-\Lambda)^\nu,\qquad
 \nu=\frac{v_+v_-}{2\beta},\notag\\[-1mm]
 \nu&=-\frac{(\mathfrak m_1+\mathfrak m_2)
 (\mathfrak m_3+\mathfrak m_4)}{2\epsilon_1\epsilon_2}.
 \label{eq:main-U1}
\end{align}
This is the standard free-boson factor in the chosen vertex normalization
~\cite{AGT,AFLT}.  It is independent of the fixed point $Y$ and should not be
confused with the horizontal Macdonald weight $u_1=e^{a\hbar}$.  Removing it
converts the naturally produced $U(2)$ answer into the pure Virasoro, or
$SU(2)$, statement:
$\mathcal F_{\rm Vir}=Z_{\rm inst}^{U(2)}/Z_{U(1)}
\equiv Z_{\rm inst}^{SU(2)}$.

\emph{From fixed-point equality to the conformal block.---}
At each level $N$ there are finitely many double partitions, and
Eq.~\eqref{eq:termwise} identifies the two summands carrying the same label
$Y$.  The generalized Cauchy identity supplies the tangent Euler denominator
and assembles the two Selberg averages into the fixed-point sum, while the
Selberg theorem supplies its four matter factors.  Their composition, rather
than the integral formula alone, is what proves AGT.  The comparison is
coefficientwise in the formal cross ratio and requires no convergence of the
full instanton series.  Since
$\mathcal F_{\rm Vir}=(1-\Lambda)^{-\nu}\cB_{\rm DF}$, one has
\begin{equation}
 [\Lambda^N]\mathcal F_{\rm Vir}
 =\sum_{k=0}^N\frac{(\nu)_k}{k!}
 [\Lambda^{N-k}]Z_{\rm inst}^{U(2)}.
 \label{eq:main-U1-convolution}
\end{equation}
Thus removing $Z_{U(1)}$ is not the deletion of selected Young diagrams; it
is a universal finite convolution of level coefficients.  All identities
are proved first at generic parameters and then extend meromorphically.  On a
resonant hypersurface the statement is their common continuation, not a new
choice of vectors in a degenerate eigenspace; no separate degeneracy
resolution is required.

\section{Conclusion}

We have proved the four-point $SU(2)$ AGT correspondence at generic
$\beta$.  The key result is the all-level factorized Selberg average 
 Theorem 1, 
Together with the generalized Cauchy identity.  Those lead to the
term-by-term equality between the DF expansion and the Nekrasov
fixed point sum.  Removing the universal $U(1)$ factor then yields the
$SU(2)$ AGT correspondence.

The residue formulation developed here also points to two natural
extensions.  The first is to generalize the corner identity and the
Selberg recursion from double partitions to $N$-tuple partitions,
which would provide a direct generalized Jack proof of the $SU(N)$
AGT correspondence.  The second is to lift the argument from the Jack
limit to generalized Macdonald polynomials and $q$-Selberg, or
Jackson, integrals.  This would lead to an analogous proof of the
Macdonald, equivalently five-dimensional $q$-deformed, AGT
correspondence
~\cite{SU3GeneralizedJack,AwataYamada5D,HS5D,Zenkevich5D,
QuantumToroidalReview}.

\section*{acknowledgments}
The authors thank Yutaka Matsuo, Jean-Emile Bourgine, Qing-jie Yuan and Qian Shen for helpful discussions.  K.Z. (Hong Zhang) is supported by a classified fund from Shanghai city.  


\ifdefined\ARXIVMERGE
\clearpage
\onecolumngrid
\begin{center}
{\large\bfseries Supplemental Material for\\[2pt]
``Proof of the AGT Conjecture at Generic $\beta$''}\\[8pt]
Le-Feng Chen and Kilar Zhang
\end{center}
\medskip
\else
\documentclass[aps,prl,preprint,superscriptaddress,nofootinbib,longbibliography]{revtex4-2}

\usepackage{amsmath,amssymb,amsthm,mathtools,bm,xcolor}
\usepackage{booktabs,array,enumitem,microtype}
\usepackage[colorlinks=true,citecolor=blue!55!black,urlcolor=blue!55!black,
linkcolor=blue!55!black]{hyperref}

\newcommand{\eps}{\varepsilon}
\newcommand{\vac}{\varnothing}
\newcommand{\dd}{\mathrm d}
\newcommand{\cR}{\mathcal R}
\newcommand{\cP}{\mathcal P}
\newcommand{\cB}{\mathcal B}
\newcommand{\cK}{\mathcal K}
\newcommand{\cL}{\mathcal L}
\newcommand{\cD}{\mathcal D}
\newcommand{\cH}{\mathcal H}
\newcommand{\ct}{\operatorname{ct}}
\newcommand{\Res}{\operatorname*{Res}}

\newtheorem{theorem}{Theorem}
\newtheorem{lemma}[theorem]{Lemma}
\theoremstyle{remark}
\newtheorem{remark}[theorem]{Remark}

\begin{document}

\title{Supplemental Material for ``Proof of the AGT Conjecture at Generic
\texorpdfstring{$\beta$}{beta}''}

\author{Le-Feng Chen}
\email{clf1@hnu.edu.cn}
\affiliation{School of Physics and Electronics,
Hunan University, Changsha 410082, China}

\author{Kilar Zhang}
\email[Corresponding author: ]{kilar@shu.edu.cn}
\affiliation{Department of Physics and Institute for Quantum Science and
Technology, Shanghai University, Shanghai 200444, China}
\affiliation{Shanghai Key Lab for Astrophysics, Shanghai 200234, China}
\affiliation{Shanghai Key Laboratory of High Temperature Superconductors,
Shanghai 200444, China}

\date{}
\maketitle
\fi

This Supplemental Material gives the normalization limits, boundary
tele-scoping identities, Selberg integration-by-parts calculation, and
coefficient-level operator checks used in the Letter.

\appendix
\setcounter{secnumdepth}{1}
\setcounter{section}{0}
\renewcommand{\thesection}{\Alph{section}}
\numberwithin{equation}{section}
\setcounter{equation}{0}
The appendices supply the calculations omitted from the Letter.  They fix the
physical and dimensionless conventions, take the generalized Macdonald Pieri
theorem to the strict-dual Jack normalization, prove the incoming row sum, and
derive the Selberg recursion from a corrected kernel-level total derivative.
The exact off-balance remainder is retained so that the scope of the
integration-by-parts identity is explicit.

\section{Conventions and the strict-dual basis}
\label{app:conventions}

Let $\epsilon_1,\epsilon_2$ be the physical equivariant parameters and set
\begin{equation}
 \beta=-\frac{\epsilon_1}{\epsilon_2}.
 \label{eq:S-physical-beta}
\end{equation}
We divide additive weights by $\epsilon_2$.  The dimensionless column and row
steps are therefore $1$ and $-\beta$, so that
\begin{equation}
 \varepsilon=1-\beta=\frac{\epsilon_1+\epsilon_2}{\epsilon_2},
 \qquad a_1=a,\quad a_2=-a,
 \qquad a=\frac{\mathfrak a}{\epsilon_2}.
 \label{eq:S-eps}
\end{equation}
For a colored box $s=(\alpha,r,c)$ define
\begin{equation}
 \operatorname{ct}(s)=c-1-\beta(r-1),\qquad
 \phi_s=a_\alpha+\operatorname{ct}(s).
 \label{eq:S-content}
\end{equation}
The corresponding physical content is
\begin{equation}
 \Phi_s=\epsilon_2\phi_s
 =\mathfrak a_\alpha+(c-1)\epsilon_2+(r-1)\epsilon_1.
 \label{eq:S-physical-content}
\end{equation}
For a double partition $Y=(\lambda,\mu)$, $A(Y)$ and $R(Y)$ are the unions of
the addable and removable corners of its two components, with the color kept
in $\phi_s$.

Let $x=(x_k)_{k\ge1}$ and $y=(y_k)_{k\ge1}$ be two power-sum alphabets.  The
two-alphabet Jack pairing used for every transpose is the product Jack
pairing of Refs.~\cite{MacdonaldBook,Ohkubo}:
\begin{equation}
 \langle x_\lambda y_\mu,x_\rho y_\sigma\rangle_\beta
 =\delta_{\lambda\rho}\delta_{\mu\sigma}z_\lambda z_\mu
 \beta^{-\ell(\lambda)-\ell(\mu)},
 \qquad
 z_\lambda=\prod_{k\ge1}k^{m_k(\lambda)}m_k(\lambda)!.
 \label{eq:S-pairing}
\end{equation}
The one-color Calogero--Sutherland operator is
\begin{align}
 \square_x={}&\frac12\sum_{r,s\ge1}rsx_{r+s}
 \partial_{x_r}\partial_{x_s}
 +\frac\beta2\sum_{r,s\ge1}(r+s)x_rx_s\partial_{x_{r+s}}
 \notag\\
 &+\frac\varepsilon2\sum_{k\ge1}k(k-1)x_k\partial_{x_k},
 \label{eq:S-square}
\end{align}
and $\square_y$ is obtained by $x\mapsto y$; the generalized two-color
Hamiltonian below is in the convention of
Refs.~\cite{MorozovSmirnov,Ohkubo}.  Define
\begin{align}
 \cH_G^\dagger={}&\square_x+a\sum_{k\ge1}kx_k\partial_{x_k}
 +\square_y-a\sum_{k\ge1}ky_k\partial_{y_k}
 \notag\\
 &+\varepsilon\sum_{k\ge1}k^2x_k\partial_{y_k},
 \label{eq:S-HG}\\
 \cD={}&\sum_{k\ge1}k(x_k\partial_{x_k}+y_k\partial_{y_k}),
 \qquad K=\cH_G^\dagger+\frac\varepsilon2\cD.
 \label{eq:S-K}
\end{align}
All transposes are taken with respect to the pairing
\eqref{eq:S-pairing}.  The monic bases $\widehat J_Y$ and $\widehat E_Y$ are
defined by the coefficientwise Jack limits of the two opposite-chamber
generalized Macdonald bases in Sec.~\ref{app:pieri-limit}; each has unit
coefficient of its leading product of ordinary monic functions.
They satisfy $K^\dagger\widehat J_Y=\kappa_Y\widehat J_Y$ and
$K\widehat E_Y=\kappa_Y\widehat E_Y$.  Set
$h_Y=\langle\widehat J_Y,\widehat E_Y\rangle$ and
$\widehat J_Y^\vee=\widehat E_Y/h_Y$.  Thus $\widehat J_Y^\vee$ is strict
dual to $\widehat J_Y$, rather than a second independently normalized basis.
The left/right triangular degeneration fixes the labeled monic pair even if
distinct double partitions share the same eigenvalue $\kappa_Y$; the
eigenvalue equation alone is not being used to choose a basis inside that
degenerate space.  In the normalization of Ref.~\cite{MorozovSmirnov},
\begin{equation}
 J_Y=N_Y\widehat J_Y,\qquad
 e_Y=\frac{\widehat J_Y^\vee}{N_Y},\qquad
 \langle J_Y,e_W\rangle=\delta_{YW}.
 \label{eq:S-dual}
\end{equation}
The more customary paper-normalized dual is
\begin{equation}
 J_Y^*=e(T_Y)e_Y,
 \qquad
 \langle J_Y,J_W^*\rangle=\delta_{YW}e(T_Y),
 \label{eq:S-Jstar}
\end{equation}
where the tangent Euler class is written explicitly in
Eq.~\eqref{eq:S-tangent} below.
The normalization is
\begin{align}
 N_{\lambda,\mu}(a,\beta)&=e_{\lambda,\mu}(2a+\beta-1),
 \label{eq:S-N}\\
 e_{\lambda,\mu}(z)
 &=\frac{(-1)^{|\lambda|}}{\beta^{|\lambda|+|\mu|}}
 \prod_{s\in\lambda}
 [z+a_\lambda(s)+1+\beta\ell_\mu(s)]
 \notag\\[-1mm]
 &\hspace{12mm}\times
 \prod_{t\in\mu}
 [z-a_\mu(t)-\beta\ell_\lambda(t)-\beta].
 \label{eq:S-e-factor}
\end{align}
Here $a_\rho(s)$ and $\ell_\rho(s)$ are the usual arm and leg lengths, extended
to a box not necessarily in $\rho$ by the same row/column formulas.  With these
conventions,
\begin{equation}
 Ke_Y=\kappa_Ye_Y,\qquad
 \kappa_Y=\sum_{s\in Y}\left(\phi_s+\frac\varepsilon2\right).
 \label{eq:S-eigen}
\end{equation}
The generalized Cauchy identity pairing the two normalizations is
\cite{MorozovSmirnov,FOS,BCS}
\begin{align}
 &\exp\!\left[\beta\sum_{k\ge1}
 \frac{P_k^{(1)}x_k+P_k^{(2)}y_k}{k}\right]
 \notag\\[-1mm]
 &\hspace{12mm}=\sum_YJ_Y(P^{(1)},P^{(2)})e_Y(x,y).
 \label{eq:S-Cauchy}
\end{align}

\section{Strict-dual Jack limit of generalized Macdonald Pieri}
\label{app:pieri-limit}

This section proves the one-box formula used in the Letter without assuming a
Selberg integral or AGT.  Set
\begin{align}
 q&=e^\hbar,\quad t=e^{\beta\hbar},\quad
 q_1=t^{-1},\quad q_2=q,\notag\\
 q_3&=(q_1q_2)^{-1}=e^{-\varepsilon\hbar},
 \qquad Q=e^{2a\hbar}.
 \label{eq:S-mult}
\end{align}
For a box $s=(r,c)$,
\begin{equation}
 \chi_s=q_1^{r-1}q_2^{c-1}=e^{\hbar\operatorname{ct}(s)}.
 \label{eq:S-chi}
\end{equation}
Thus every multiplicative difference is converted into an additive one through
$1-e^{\hbar A}=-\hbar A+O(\hbar^2)$.

Ohkubo's coefficientwise degeneration~\cite{Ohkubo} defines the two monic
generalized Jack bases used below:
\begin{equation}
 \widehat E_{\lambda,\mu}(x,y)=\lim_{\hbar\to0}
 P_{\lambda,\mu}(x,y\mid Q),\qquad
 \widehat J_{\lambda,\mu}(x,y)=\lim_{\hbar\to0}
 P_{\mu,\lambda}(y,x\mid Q^{-1}).
 \label{eq:S-two-Jack-limits}
\end{equation}

For an ordinary partition $\rho$, define the multiplicative corner functions
\begin{equation}
 \mathcal Y_\rho(z)=
 \frac{\prod_{b\in A(\rho)}(1-\chi_b/z)}
 {\prod_{r\in R(\rho)}(1-\chi_r/(q_3z))},
 \qquad
 \Psi_\rho(z)=\frac{\mathcal Y_\rho(q_3^{-1}z)}{\mathcal Y_\rho(z)}.
 \label{eq:S-Ymult}
\end{equation}
The two-component generalized Macdonald $e_1$-Pieri theorem of
Ref.~\cite{BCS} reads
\begin{align}
 &(x_1+y_1)P_{\lambda,\mu}(x,y\mid Q)
 \notag\\
 &=\sum_{x\in A(\lambda)}\psi_\lambda(x)\Psi_\mu(Q\chi_x)
 P_{\lambda+x,\mu}(x,y\mid Q)
 \notag\\
 &\quad+\sum_{x\in A(\mu)}\psi_\mu(x)
 P_{\lambda,\mu+x}(x,y\mid Q).
 \label{eq:S-MacPieri}
\end{align}
The factor $\Psi_\mu$ makes Eq.~\eqref{eq:S-MacPieri} genuinely two-colored;
it is not a juxtaposition of two ordinary Pieri formulas.
For an addable box $x=(i,j)$ of an ordinary partition $\rho$, the ordinary
coefficient appearing here is
\begin{equation}
 \psi_\rho(i,j)=\prod_{i'=1}^{i-1}
 \frac{(1-q^{\rho_{i'}-j+1}t^{i-i'-1})
       (1-q^{\rho_{i'}-j}t^{i-i'+1})}
      {(1-q^{\rho_{i'}-j+1}t^{i-i'})
       (1-q^{\rho_{i'}-j}t^{i-i'})}.
 \label{eq:S-ordinary-MacPieri}
\end{equation}

For an ordinary monic Macdonald polynomial $P_\rho$, put
\begin{align}
 b_\rho(q,t)&=\prod_{s\in\rho}
 \frac{1-q^{a_\rho(s)}t^{\ell_\rho(s)+1}}
 {1-q^{a_\rho(s)+1}t^{\ell_\rho(s)}},
 \notag\\
 p_k(\mathrm{sp}_\vac)&=\frac{q_1^k}{1-q_1^k},
 \qquad
 \widetilde b_\rho=b_\rho P_\rho(\mathrm{sp}_\vac)^2.
 \label{eq:S-spherical-data}
\end{align}
The two spherical factors paired by the Cauchy kernel are, explicitly,
\begin{align}
 \widetilde P_Y^{(+)}(x,y\mid Q)
 &=\frac{P_{\lambda,\mu}(x,y\mid Q)}
 {P_\lambda(\mathrm{sp}_\vac)P_\mu(\mathrm{sp}_\vac)},
 \notag\\
 \widetilde P_Y^{(-)}(a',b'\mid Q)
 &=\frac{P_{\mu,\lambda}(a',b'\mid Q^{-1})}
 {P_\mu(\mathrm{sp}_\vac)P_\lambda(\mathrm{sp}_\vac)}.
 \label{eq:S-two-spherical-factors}
\end{align}
Thus the first factor, not an implicitly reversed copy of it, is the one in
the Pieri rule \eqref{eq:S-MacPieri}.  Remove its homogeneous zero by
\begin{equation}
 \mathsf P_Y^{(\hbar)}=(\beta\hbar)^{-|Y|}
 \widetilde P_Y^{(+)}.
 \label{eq:S-Prescale}
\end{equation}
Their relative strict-dual normalization follows from the generalized
Macdonald Cauchy kernel~\cite{FOS,BCS},
\begin{align}
 \Pi_Z={}&\sum_{\lambda,\mu}\widetilde b_\lambda\widetilde b_\mu
 \widetilde P_Y^{(+)}(x,y\mid Q)
 \widetilde P_Y^{(-)}(a',b'\mid Q)
 \notag\\
 ={}&\exp\!\left[\sum_{k\ge1}\frac1k\frac{1-t^k}{1-q^k}
 \left\{x_kb'_k+y_ka'_k\right.\right.\notag\\[-1mm]
 &\hspace{34mm}\left.\left.+(1-t^kq^{-k})x_ka'_k\right\}\right].
 \label{eq:S-MacCauchy}
\end{align}
Since $(1-t^k)/(1-q^k)\to\beta$ and
$1-t^kq^{-k}=O(\hbar)$, Eq.~\eqref{eq:S-MacCauchy} tends to
Eq.~\eqref{eq:S-Cauchy}.  After exchanging $(a',b')$ to undo the displayed
component reversal, $\widetilde P^{(-)}$ tends to $\widehat J_Y$, while the
Pieri factor $\widetilde P^{(+)}$ tends to its Cauchy-dual partner.  If
\begin{equation}
 C_Y^{(\hbar)}=\widetilde b_\lambda\widetilde b_\mu
 (\beta\hbar)^{2|Y|},
 \label{eq:S-C}
\end{equation}
then coefficient comparison in the Cauchy kernel fixes
\begin{equation}
 \lim_{\hbar\to0}C_Y^{(\hbar)}\mathsf P_Y^{(\hbar)}
 =\widehat J_Y^\vee.
 \label{eq:S-duallimit}
\end{equation}

The paper normalization $N_Y$ is already fixed by
Eqs.~\eqref{eq:S-N}--\eqref{eq:S-e-factor}.  Let $N_Y^{(\hbar)}$ be its
finite-$\hbar$ generalized-Macdonald deformation in the same triangular
chamber, normalized by
$\lim_{\hbar\to0}N_Y^{(\hbar)}=N_Y$.  Define the rescaled Pieri basis
\begin{equation}
 \mathsf e_Y^{(\hbar)}=
 \frac{C_Y^{(\hbar)}}{N_Y^{(\hbar)}}\mathsf P_Y^{(\hbar)}.
 \label{eq:S-eh}
\end{equation}
Equation~\eqref{eq:S-duallimit} immediately implies
$\mathsf e_Y^{(\hbar)}\to e_Y$.

Write the spherical form of Eq.~\eqref{eq:S-MacPieri} as
\begin{equation}
 M\widetilde P_W=\sum_{Y=W+x}\widetilde c_{Y/W}\widetilde P_Y,
 \qquad M=x_1+y_1.
 \label{eq:S-sphPieri}
\end{equation}
Since $|Y|=|W|+1$, substitution of Eqs.~\eqref{eq:S-Prescale} and
\eqref{eq:S-eh} gives the exact finite-$\hbar$ identity
\begin{equation}
 M\mathsf e_W^{(\hbar)}=
 \sum_{Y=W+x}(\beta\hbar)\widetilde c_{Y/W}
 \frac{C_W^{(\hbar)}}{C_Y^{(\hbar)}}
 \frac{N_Y^{(\hbar)}}{N_W^{(\hbar)}}
 \mathsf e_Y^{(\hbar)}.
 \label{eq:S-extract}
\end{equation}
This equation is the precise extraction of each final-state coefficient from
the weighted Pieri sum.  Linear independence allows every coefficient to be
limited separately:
\begin{equation}
 M_{Y/W}=\lim_{\hbar\to0}(\beta\hbar)\widetilde c_{Y/W}
 \frac{C_W^{(\hbar)}}{C_Y^{(\hbar)}}
 \frac{N_Y^{(\hbar)}}{N_W^{(\hbar)}}.
 \label{eq:S-coefflimit}
\end{equation}

For completeness, we now evaluate this limit in corner form.  Introduce
\begin{equation}
 y_\rho(d)=
 \frac{\prod_{b\in A(\rho)}[d-\operatorname{ct}(b)]}
 {\prod_{r\in R(\rho)}[d-\operatorname{ct}(r)-\varepsilon]}.
 \label{eq:S-yadd}
\end{equation}
Horizontal and vertical neighbor factors telescope, giving the equivalent
box-product formula
\begin{equation}
 y_\rho(d)=d\prod_{s\in\rho}
 \frac{[d-\operatorname{ct}(s)-1]
       [d-\operatorname{ct}(s)+\beta]}
      {[d-\operatorname{ct}(s)]
       [d-\operatorname{ct}(s)-\varepsilon]}.
 \label{eq:S-y-box-product}
\end{equation}
Then
\begin{equation}
 \mathcal Y_\rho(e^{\hbar d})=\hbar y_\rho(d)+O(\hbar^2),
 \qquad
 \Psi_\rho(e^{\hbar d})\longrightarrow
 \frac{y_\rho(d+\varepsilon)}{y_\rho(d)}.
 \label{eq:S-Psilimit}
\end{equation}
To obtain the normalization ratios, divide the product in
Eq.~\eqref{eq:S-e-factor} after and before adding $x$.  Factors outside the
row and column meeting $x$ cancel.  The remaining horizontal ratios telescope
to the factors with shifts $-1$ and $0$ in
Eq.~\eqref{eq:S-y-box-product}, while the vertical ratios telescope to the
shifts $+\beta$ and $-\varepsilon$.  The uncancelled corner factor supplies the
initial $d$.  This gives, without a division by a diagram-dependent sum,
\begin{align}
 \frac{N_{\lambda+x,\mu}}{N_{\lambda,\mu}}
 &=-\frac{1}{\beta}y_\mu(2a+\operatorname{ct}(x)),
 &&x\in A(\lambda),
 \label{eq:S-Nratio1}\\
 \frac{N_{\lambda,\mu+x}}{N_{\lambda,\mu}}
 &=-\frac{1}{\beta}y_\lambda(-2a+\operatorname{ct}(x)+\varepsilon),
 &&x\in A(\mu).
 \label{eq:S-Nratio2}
\end{align}
The ordinary strict-dual Pieri factor can be obtained before taking the limit.
After combining the ordinary coefficient in
Eq.~\eqref{eq:S-ordinary-MacPieri} with its spherical and Cauchy-dual
normalization ratios, the one-color factor is
\begin{equation*}
 \mathsf D_\rho^{(\hbar)}(x)=
 \frac{1-q}{1-t}
 \frac{\displaystyle\prod_{b\in A(\rho)\setminus\{x\}}
 (1-q_3\chi_b/\chi_x)}
 {\displaystyle\prod_{r\in R(\rho)}(1-\chi_r/\chi_x)}.
\end{equation*}
This corner form is the telescoped version of the row product in
Eq.~\eqref{eq:S-ordinary-MacPieri}.  Its additive limit is factor by factor:
\begin{align*}
 \frac{1-q}{1-t}&\longrightarrow\frac1\beta,\\
 1-q_3\frac{\chi_b}{\chi_x}
 &=\hbar[\operatorname{ct}(x)-\operatorname{ct}(b)+\varepsilon]
   +O(\hbar^2),\\
 1-\frac{\chi_r}{\chi_x}
 &=\hbar[\operatorname{ct}(x)-\operatorname{ct}(r)]
   +O(\hbar^2).
\end{align*}
Because $|A(\rho)|-1=|R(\rho)|$, the powers of $\hbar$ cancel, and hence
\begin{equation}
 D_\rho(x)={1\over\beta}
 {\displaystyle\prod_{b\in A(\rho)\setminus\{x\}}
 [\operatorname{ct}(x)-\operatorname{ct}(b)+\varepsilon]
 \over
 \displaystyle\prod_{r\in R(\rho)}
 [\operatorname{ct}(x)-\operatorname{ct}(r)]}.
 \label{eq:S-D}
\end{equation}
To display the remaining limit without hiding a normalization, define
\begin{equation*}
 \mathsf L_{Y/W}^{(\hbar)}=
 (\beta\hbar)\widetilde c_{Y/W}
 \frac{C_W^{(\hbar)}}{C_Y^{(\hbar)}}.
\end{equation*}
For $x\in A(\lambda)$, put
$d_+=2a+\operatorname{ct}(x)$.  Equations~\eqref{eq:S-Psilimit} and
\eqref{eq:S-D} give
\begin{equation*}
 \mathsf L_{(\lambda+x,\mu)/(\lambda,\mu)}^{(\hbar)}
 \longrightarrow
 D_\lambda(x)\frac{y_\mu(d_++\varepsilon)}{y_\mu(d_+)}.
\end{equation*}
The remaining paper-normalization ratio is
$-y_\mu(d_+)/\beta$ by Eq.~\eqref{eq:S-Nratio1}.  Therefore the factor
$y_\mu(d_+)$ cancels and
\begin{equation}
 M_{(\lambda+x,\mu)/(\lambda,\mu)}
 =-{1\over\beta}D_\lambda(x)y_\mu(d_++\varepsilon).
 \label{eq:S-color1}
\end{equation}
For $x\in A(\mu)$, put
$d_-=-2a+\operatorname{ct}(x)+\varepsilon$.  The second branch of the
generalized Macdonald theorem contains no $\Psi$ factor, so
\begin{equation*}
 \mathsf L_{(\lambda,\mu+x)/(\lambda,\mu)}^{(\hbar)}
 \longrightarrow D_\mu(x).
\end{equation*}
Multiplication by the ratio $-y_\lambda(d_-)/\beta$ from
Eq.~\eqref{eq:S-Nratio2} gives
\begin{equation}
 M_{(\lambda,\mu+x)/(\lambda,\mu)}
 =-{1\over\beta}D_\mu(x)
 y_\lambda(d_-).
 \label{eq:S-color2}
\end{equation}
It remains only to combine the two colors.  If $x\in A(\lambda)$, then
$\phi_x=a+\operatorname{ct}(x)$ and direct substitution into
Eqs.~\eqref{eq:S-yadd} and \eqref{eq:S-D} gives
\begin{align*}
 &D_\lambda(x)y_\mu(d_++\varepsilon)\\
 &\quad={1\over\beta}
 \frac{\displaystyle
  \prod_{b\in A(\lambda)\setminus\{x\}}
  [\operatorname{ct}(x)-\operatorname{ct}(b)+\varepsilon]
  \prod_{b\in A(\mu)}
  [2a+\operatorname{ct}(x)-\operatorname{ct}(b)+\varepsilon]}
 {\displaystyle
  \prod_{r\in R(\lambda)}
  [\operatorname{ct}(x)-\operatorname{ct}(r)]
  \prod_{r\in R(\mu)}
  [2a+\operatorname{ct}(x)-\operatorname{ct}(r)]}\\
 &\quad={1\over\beta}
 \frac{\displaystyle\prod_{b\in A(W)\setminus\{x\}}
  (\phi_x-\phi_b+\varepsilon)}
 {\displaystyle\prod_{r\in R(W)}(\phi_x-\phi_r)}.
\end{align*}
If $x\in A(\mu)$, the same calculation with
$\phi_x=-a+\operatorname{ct}(x)$ gives
\begin{equation*}
 D_\mu(x)y_\lambda(d_-)
 ={1\over\beta}
 \frac{\displaystyle\prod_{b\in A(W)\setminus\{x\}}
  (\phi_x-\phi_b+\varepsilon)}
 {\displaystyle\prod_{r\in R(W)}(\phi_x-\phi_r)}.
\end{equation*}
Thus Eqs.~\eqref{eq:S-color1} and \eqref{eq:S-color2} combine into
\begin{equation}
 \boxed{
 M_{W+x/W}=-{1\over\beta^2}
 {\displaystyle\prod_{b\in A(W)\setminus\{x\}}
 (\phi_x-\phi_b+\varepsilon)
 \over
 \displaystyle\prod_{r\in R(W)}(\phi_x-\phi_r)}.}
 \label{eq:S-JackPieri}
\end{equation}
Every equation in this section is all-level.  Low-level polynomials are not
used to infer Eq.~\eqref{eq:S-JackPieri}.

\section{Residue proof of the incoming row sum}
\label{app:row-sum}

Fix a nonempty final state $Y$.  For each $x\in R(Y)$ set $W=Y-x$ and
write
\begin{equation}
 U_x(W)=
 {\displaystyle\prod_{b\in A(W)\setminus\{x\}}
 (\phi_x-\phi_b+\varepsilon)
 \over
 \displaystyle\prod_{r\in R(W)}(\phi_x-\phi_r)}.
 \label{eq:S-U}
\end{equation}
We use the standard fixed-point $Y$-function box/corner identity
\cite{BCS}; Eq.~\eqref{eq:S-Pbox} below is its additive Jack-limit form.
For completeness we derive that form explicitly in our content conventions.
Define a single rational function whose poles label all the parents $W$:
\begin{equation}
 \begin{aligned}
 \cP_Y(z)&=\frac{\mathsf A_Y(z)}{\mathsf R_Y(z-\varepsilon)},\\
 \mathsf A_Y(z)&=\prod_{b\in A(Y)}(z-\phi_b),\\
 \mathsf R_Y(z)&=\prod_{r\in R(Y)}(z-\phi_r).
 \end{aligned}
 \label{eq:S-P}
\end{equation}
The denominator forces a simple pole at $z=\phi_x+\varepsilon$ for every
$x\in R(Y)$.  Its residue is
\begin{equation}
 \mathop{\rm Res}_{z=\phi_x+\varepsilon}\cP_Y(z)\dd z
 =
 {\displaystyle\prod_{b\in A(Y)}(\phi_x-\phi_b+\varepsilon)
 \over
 \displaystyle\prod_{r\in R(Y)\setminus\{x\}}(\phi_x-\phi_r)}.
 \label{eq:S-resraw}
\end{equation}
We first prove the boundary-to-box identity rather than assuming it.  Consider
one component $\rho=(\rho_1,\ldots,\rho_\ell)$ of color $\alpha$ and put
$u_i=z-a_\alpha+\beta(i-1)$.  For the box $(i,j)$ one has
$z-\phi_{i,j}=u_i-j+1$, and the product along row $i$ telescopes as
\begin{align*}
 &\prod_{j=1}^{\rho_i}
 \frac{(z-\phi_{i,j}-1)(z-\phi_{i,j}+\beta)}
 {(z-\phi_{i,j})(z-\phi_{i,j}-\varepsilon)}\\
 &\quad=
 \left[\prod_{j=1}^{\rho_i}\frac{u_i-j}{u_i-j+1}\right]
 \left[\prod_{j=1}^{\rho_i}
 \frac{u_i-j+1+\beta}{u_i-j+\beta}\right]\\
 &\quad=\frac{(u_i-\rho_i)(u_i+\beta)}
 {u_i(u_i-\rho_i+\beta)}.
\end{align*}
Since $u_{i+1}=u_i+\beta$, multiplication over all rows, including the
vacuum factor $z-a_\alpha=u_1$, gives
\begin{equation*}
 (z-a_\alpha)\prod_{s\in\rho}
 \frac{(z-\phi_s-1)(z-\phi_s+\beta)}
 {(z-\phi_s)(z-\phi_s-\varepsilon)}
 =u_{\ell+1}\prod_{i=1}^{\ell}
 \frac{u_i-\rho_i}{u_{i+1}-\rho_i}.
\end{equation*}
Here $u_i-\rho_i$ is the factor $z-\phi_b$ for the candidate addable box
$b=(i,\rho_i+1)$, $u_{\ell+1}$ is the factor for the new-row addable box,
and $u_{i+1}-\rho_i$ is $z-\phi_r-\varepsilon$ for the candidate removable box
$r=(i,\rho_i)$.  If $\rho_i=\rho_{i+1}$, the corresponding denominator and
next numerator coincide and cancel.  Thus precisely the genuine addable and
removable corners remain:
\begin{equation*}
 (z-a_\alpha)\prod_{s\in\rho}
 \frac{(z-\phi_s-1)(z-\phi_s+\beta)}
 {(z-\phi_s)(z-\phi_s-\varepsilon)}
 =\frac{\prod_{b\in A(\rho)}(z-\phi_b)}
 {\prod_{r\in R(\rho)}(z-\phi_r-\varepsilon)}.
\end{equation*}
Multiplying this identity for the two colors gives
\begin{equation}
 \cP_Y(z)=\prod_{\alpha=1}^2(z-a_\alpha)
 \prod_{s\in Y}
 { (z-\phi_s-1)(z-\phi_s+\beta)
 \over (z-\phi_s)(z-\phi_s-\varepsilon)}.
 \label{eq:S-Pbox}
\end{equation}
This identity also gives the residue weight without a local case split.  If
$Y=W+x$, then $\cP_Y(z)=\cP_W(z)f_x(z)$ with
\begin{equation*}
 f_x(z)=\frac{(z-\phi_x-1)(z-\phi_x+\beta)}
 {(z-\phi_x)(z-\phi_x-\varepsilon)}.
\end{equation*}
At $z_x=\phi_x+\varepsilon$,
\begin{equation*}
 \mathop{\rm Res}_{z=z_x}f_x(z)\dd z
 =\frac{(\varepsilon-1)(\varepsilon+\beta)}{\varepsilon}
 =-\frac\beta\varepsilon.
\end{equation*}
On the other hand, the $b=x$ factor in the numerator of $\cP_W(z_x)$ is
$\varepsilon$, so
\begin{equation*}
 \cP_W(z_x)=\varepsilon
 {\prod_{b\in A(W)\setminus\{x\}}(\phi_x-\phi_b+\varepsilon)
 \over\prod_{r\in R(W)}(\phi_x-\phi_r)}
 =\varepsilon U_x(W).
\end{equation*}
Their product proves
\begin{equation}
 {\prod_{b\in A(Y)}(\phi_x-\phi_b+\varepsilon)
 \over\prod_{r\in R(Y)\setminus\{x\}}(\phi_x-\phi_r)}
 =-\beta U_x(W).
 \label{eq:S-cornerchange}
\end{equation}
The derivation is made for generic $\varepsilon\ne0$; the final rational identity
extends to $\varepsilon=0$ by meromorphic continuation.

The single-box factor has the large-$z$ expansion
\begin{equation}
 \begin{aligned}
 &\frac{(z-\phi-1)(z-\phi+\beta)}
 {(z-\phi)(z-\phi-\varepsilon)}\\
 &\qquad=1-\frac{\beta}{z^2}
 -\frac{\beta(2\phi+\varepsilon)}{z^3}
 +O(z^{-4}).
 \end{aligned}
 \label{eq:S-oneboxexp}
\end{equation}
Because $a_1+a_2=0$,
$\prod_\alpha(z-a_\alpha)=z^2-a^2$.  Multiplying
Eq.~\eqref{eq:S-oneboxexp} over all boxes gives
\begin{align}
 \cP_Y(z)={}&z^2-(a^2+\beta|Y|)\notag\\
 &-\frac{\beta\sum_{s\in Y}(2\phi_s+\varepsilon)}{z}
 +O(z^{-2}).
 \label{eq:S-infinity}
\end{align}
For a rational one-form,
$\operatorname{Res}_{z=\infty}\cP_Y(z)\dd z
=-[z^{-1}]\cP_Y(z)$; hence the sum of the finite residues equals the
coefficient of $z^{-1}$ in the large-$z$ expansion.
Equations~\eqref{eq:S-cornerchange} and \eqref{eq:S-infinity} therefore imply
\begin{equation}
 \sum_{x\in R(Y)}U_x(Y-x)
 =\sum_{s\in Y}(2\phi_s+\varepsilon)=2\kappa_Y.
 \label{eq:S-Usum}
\end{equation}
Using Eq.~\eqref{eq:S-JackPieri} for every incoming arrow proves
\begin{equation}
 \boxed{\displaystyle
 \sum_{W=Y-x}M_{Y/W}=-{2\over\beta^2}\kappa_Y.}
 \label{eq:S-rowsum}
\end{equation}

As a symbolic check, take $Y=([1],\vac)$.  Its addable contents are
$a_1+1$, $a_1-\beta$, and $a_2$, while its removable content is
$a_1$.  Thus
\begin{equation}
 \cP_Y(z)=
 {(z-a_1-1)(z-a_1+\beta)(z-a_2)
 \over z-a_1-\varepsilon},
 \label{eq:S-oneboxP}
\end{equation}
and the residue at $a_1+\varepsilon$ is
$-\beta(a_1-a_2+\varepsilon)$, exactly
$-\beta U_x(\vac,\vac)$.

\section{Generating function and mass operator}
\label{app:mass-operator}

Let
\begin{align}
 p_k(t)&=\sum_{i=1}^nt_i^k,\qquad R_k=y_k-x_k,\notag\\
 a&=-\beta n-\frac{u+v+\varepsilon}{2},
 \label{eq:S-Selbergpars}
\end{align}
and define the normalized Selberg average by
\begin{equation}
 \langle f\rangle_S={1\over Z_S}\int_{[0,1]^n}f(t)
 \prod_{i<j}|t_i-t_j|^{2\beta}
 \prod_i t_i^u(1-t_i)^v\dd t_i.
 \label{eq:S-average}
\end{equation}
Introduce
\begin{equation}
 \Psi(x,y)=\left\langle
 \exp\!\left[\beta\sum_{k\ge1}
 {(-p_k-v/\beta)x_k+p_ky_k\over k}\right]
 \right\rangle_S.
 \label{eq:S-Psi}
\end{equation}
Equation~\eqref{eq:S-Cauchy} gives the exact expansion
\begin{equation}
 \Psi=\sum_YF_Y^{\rm Sel}e_Y,\qquad
 F_Y^{\rm Sel}=\left\langle
 J_Y\left(a,-p_k-{v\over\beta},p_k\right)\right\rangle_S.
 \label{eq:S-Psiexpand}
\end{equation}

Set $M=x_1+y_1$ and
\begin{equation}
 B=[K,M],\qquad C=[K,B].
 \label{eq:S-BC}
\end{equation}
A direct use of Eqs.~\eqref{eq:S-square}--\eqref{eq:S-K} yields
\begin{align}
 B={}&\sum_{k\ge1}k(x_{k+1}\partial_{x_k}+y_{k+1}\partial_{y_k})
 \notag\\[-1mm]
 &+\left(a+\frac{3\varepsilon}{2}\right)x_1
 +\left(-a+\frac{\varepsilon}{2}\right)y_1.
 \label{eq:S-B}
\end{align}
Define
\begin{equation}
 m_1={u-v+\varepsilon\over2},\qquad
 m_2={-u-v-\varepsilon\over2},
 \label{eq:S-masses}
\end{equation}
and
\begin{equation}
 \cR_{m_1,m_2}={1\over\beta^2}
 \left[C-(v+\varepsilon)B
 -{(u-v)(u+v+2\varepsilon)\over4}M\right].
 \label{eq:S-R}
\end{equation}
If $Y=W+x$, Eq.~\eqref{eq:S-eigen} gives
$\kappa_Y-\kappa_W=\phi_x+\varepsilon/2$.  Hence
\begin{align}
 B_{Y/W}&=(\phi_x+\varepsilon/2)M_{Y/W},
 &C_{Y/W}&=(\phi_x+\varepsilon/2)^2M_{Y/W}.
 \label{eq:S-BCmatrix}
\end{align}
The elementary identity
\begin{align}
 &\left(\phi+{\varepsilon\over2}\right)^2
 -(v+\varepsilon)\left(\phi+{\varepsilon\over2}\right)
 -{(u-v)(u+v+2\varepsilon)\over4}
 \notag\\[-1mm]
 &\hspace{45mm}=(\phi+m_1)(\phi+m_2)
 \label{eq:S-quadratic}
\end{align}
therefore gives
\begin{equation}
 (\cR_{m_1,m_2})_{Y/W}
 ={(\phi_x+m_1)(\phi_x+m_2)\over\beta^2}M_{Y/W}.
 \label{eq:S-Rmatrix}
\end{equation}

\section{Corrected all-order Selberg total derivative}
\label{app:total-derivative}

We prove the operator identity
\begin{equation}
 K\Psi={\beta^2\over2}\cR_{m_1,m_2}\Psi.
 \label{eq:S-intertwining}
\end{equation}
Before averaging, write
\begin{equation}
 \cK(x,y;t)=\exp\!\left[-v\sum_{k\ge1}{x_k\over k}
 +\beta\sum_{k\ge1}{p_k(t)R_k\over k}\right].
 \label{eq:S-kernel}
\end{equation}
For this calculation only, release the balance condition temporarily and set
\begin{equation}
 a_0=-\beta n-\frac{u+v+\varepsilon}{2},
 \qquad \delta=a-a_0.
 \label{eq:S-a0-delta}
\end{equation}
Then
\begin{equation}
 \cK^{-1}\partial_{x_k}\cK=-{v+\beta p_k\over k},
 \qquad
 \cK^{-1}\partial_{y_k}\cK={\beta p_k\over k}.
 \label{eq:S-kernelder}
\end{equation}
Let $\Delta_S(t)$ be the unnormalized weight in
Eq.~\eqref{eq:S-average}, and introduce
\begin{equation}
 S(t)=\sum_{k\ge1}(x_k+y_k)t^k,\qquad
 f_i=-{\beta\over2}(1-t_i)S(t_i).
 \label{eq:S-flux}
\end{equation}
The exact statement, with $a,u,v,n$ initially independent, is
\begin{align}
 &\Delta_S\left(K-\frac{\beta^2}{2}\cR_{m_1,m_2}\right)\cK
 -\sum_{i=1}^n\partial_{t_i}(f_i\Delta_S\cK)
 \notag\\
 &\quad=\delta\,\Delta_S\cK
 \left[
 \beta\sum_{k\ge1}p_k(S_{k+1}-S_k)
 -\frac{v+a+a_0}{2}S_1
 \right].
 \label{eq:S-master-totalder}
\end{align}
Consequently the pure total-derivative identity is
\begin{equation}
 \boxed{
 \Delta_S\left(K-\frac{\beta^2}{2}\cR_{m_1,m_2}\right)\cK
 =\sum_{i=1}^n\partial_{t_i}(f_i\Delta_S\cK)}
 \qquad (a=a_0).
 \label{eq:S-totalder}
\end{equation}
Thus Eq.~\eqref{eq:S-totalder} would be false if $a$ were treated as an
independent parameter.  We now derive both the balanced identity and the
off-balance remainder.
Expand
\begin{equation}
 f(t)=\sum_{\ell\ge1}f_\ell t^\ell,\qquad
 f_\ell=-{\beta\over2}(S_\ell-S_{\ell-1}),\qquad S_0=0,
 \label{eq:S-fell}
\end{equation}
where $S_k=x_k+y_k$.  Dividing the right-hand side of
Eq.~\eqref{eq:S-totalder} by $\Delta_S\cK$ gives
\begin{align}
 \mathcal T_f={}&\sum_{\ell\ge1}(\ell+u)f_\ell p_{\ell-1}
 +\beta\sum_{\ell\ge1}f_\ell
 \sum_{r=0}^{\ell-1}p_rp_{\ell-1-r}
 \notag\\
 &-\beta\sum_{\ell\ge1}\ell f_\ell p_{\ell-1}
 \notag\\
 &+{\beta v\over2}\sum_{k\ge1}S_kp_k
 +\beta\sum_{\ell,k\ge1}f_\ell R_kp_{k+\ell-1},
 \label{eq:S-Tf}
\end{align}
where $p_0=n$.  The Vandermonde term uses
\begin{equation}
 \sum_{i<j}{t_i^\ell-t_j^\ell\over t_i-t_j}
 ={1\over2}\left(\sum_{r=0}^{\ell-1}p_rp_{\ell-1-r}
 -\ell p_{\ell-1}\right).
 \label{eq:S-vander}
\end{equation}
Independently, rewrite $K$ in the variables
$S_k=x_k+y_k$ and $R_k=y_k-x_k$.  Direct substitution gives
\begin{align}
 K={}&\frac12\sum_{r,s\ge1}rs\Bigl[
 S_{r+s}(\partial_{S_r}\partial_{S_s}
 +\partial_{R_r}\partial_{R_s})
 +2R_{r+s}\partial_{S_r}\partial_{R_s}\Bigr]
 \notag\\
 &+\frac\beta4\sum_{r,s\ge1}(r+s)\Bigl[
 (S_rS_s+R_rR_s)\partial_{S_{r+s}}
 +(S_rR_s+R_rS_s)\partial_{R_{r+s}}\Bigr]
 \notag\\
 &+\sum_{k\ge1}\left[
 \varepsilon k^2S_k-\left(ak+\frac\varepsilon2k^2\right)R_k
 \right]\partial_{S_k}
 +\sum_{k\ge1}\left(\frac\varepsilon2k^2-ak\right)
 S_k\partial_{R_k}.
 \label{eq:S-K-SR}
\end{align}
The kernel takes the form
\begin{equation}
 \cK=\exp\left[-\frac v2\sum_{k\ge1}\frac{S_k}{k}
 +\sum_{k\ge1}\frac{\beta p_k+v/2}{k}R_k\right],
 \label{eq:S-kernel-SR}
\end{equation}
so conjugation is exactly
\begin{align}
 \cK^{-1}\partial_{S_k}\cK
 &=\partial_{S_k}-\frac{v}{2k},\notag\\
 \cK^{-1}\partial_{R_k}\cK
 &=\partial_{R_k}+\frac{\beta p_k+v/2}{k}.
 \label{eq:S-conjugation-shifts}
\end{align}
Define
\begin{equation}
 \cL_a:=K-\frac{\beta^2}{2}\cR_{m_1,m_2}
 =K-\frac12C+\frac{v+\varepsilon}{2}B
 +\frac{(u-v)(u+v+2\varepsilon)}8M.
 \label{eq:S-L-a}
\end{equation}
Substitution of Eqs.~\eqref{eq:S-K-SR}, \eqref{eq:S-B}, and
\eqref{eq:S-conjugation-shifts}, followed by action on $1$, gives on
$a=a_0$ the following four summed monomial families:
\begin{align}
 (S_\ell-S_{\ell-1})p_{\ell-1}:&
 &&-{\beta\over2}(u+\varepsilon\ell),
 \notag\\
 (S_\ell-S_{\ell-1})\sum_{r=0}^{\ell-1}p_rp_{\ell-1-r}:&
 &&-{\beta^2\over2},
 \notag\\
 (S_\ell-S_{\ell-1})R_kp_{k+\ell-1}:&
 &&-{\beta^2\over2},
 \notag\\
 S_kp_k:& &&{\beta v\over2}.
 \label{eq:S-structures}
\end{align}
Substitution of Eq.~\eqref{eq:S-fell} into Eq.~\eqref{eq:S-Tf}
reproduces every line of Eq.~\eqref{eq:S-structures}.  For example, the
linear coefficient is
\begin{equation}
 [\ell+u-\beta\ell]f_\ell p_{\ell-1}
 =-{\beta\over2}(u+\varepsilon\ell)
 (S_\ell-S_{\ell-1})p_{\ell-1}.
 \label{eq:S-linearcheck}
\end{equation}
The four displayed lines are a grouping of the full sums, not a claim of
algebraic independence after $p_0=n$ is inserted.  The explicit operator
ledger in Appendix~\ref{app:operator-ledger} eliminates every derivative and
reproduces all four sums.  Comparing them with Eq.~\eqref{eq:S-Tf} proves
Eq.~\eqref{eq:S-totalder} on $a=a_0$ with no level cutoff.

It remains to verify the claimed off-balance term.  Write
\begin{equation}
 K(a)=K(0)+aA,
 \qquad
 A=-\sum_{k\ge1}k
 (R_k\partial_{S_k}+S_k\partial_{R_k}),
 \label{eq:S-A}
\end{equation}
and
\begin{equation}
 T_\times=\sum_{k\ge1}k
 (R_{k+1}\partial_{S_k}+S_{k+1}\partial_{R_k}).
 \label{eq:S-Tcross}
\end{equation}
Because $\partial_aB=[A,M]=-R_1$ and $C=[K,B]$, direct differentiation of
Eq.~\eqref{eq:S-L-a} gives
\begin{equation}
 \frac{\partial\cL_a}{\partial a}
 =A+T_\times-\frac v2R_1-aS_1.
 \label{eq:S-dL-da}
\end{equation}
Integrating in $a$ from $a_0$ to $a$ yields
\begin{equation}
 \cL_a-\cL_{a_0}
 =(a-a_0)\left[
 A+T_\times-\frac v2R_1-\frac{a+a_0}{2}S_1
 \right].
 \label{eq:S-L-difference}
\end{equation}
Using Eq.~\eqref{eq:S-conjugation-shifts},
\begin{align}
 \cK^{-1}A\cK\,1
 &=\sum_{k\ge1}\left[
 \frac v2R_k-\left(\beta p_k+\frac v2\right)S_k\right],
 \label{eq:S-A-kernel}\\
 \cK^{-1}T_\times\cK\,1
 &=\sum_{k\ge1}\left[
 -\frac v2R_{k+1}
 +\left(\beta p_k+\frac v2\right)S_{k+1}\right].
 \label{eq:S-Tcross-kernel}
\end{align}
The $R$ terms telescope and cancel the explicit $-vR_1/2$; the constant
$S$ terms telescope to $-vS_1/2$.  Hence
\begin{align}
 \cK^{-1}(\cL_a-\cL_{a_0})\cK\,1
 &=(a-a_0)\left[
 \beta\sum_{k\ge1}p_k(S_{k+1}-S_k)
 -\frac{v+a+a_0}{2}S_1\right],
 \label{eq:S-off-balance-scalar}
\end{align}
which proves Eq.~\eqref{eq:S-master-totalder}.  In particular, its first
obstruction contains
\begin{equation}
 -(a-a_0)S_1\left[\beta p_1+\frac{v+a+a_0}{2}\right],
 \label{eq:S-first-obstruction}
\end{equation}
and is generically nonzero away from charge balance.

For $\operatorname{Re}u>-1$, $\operatorname{Re}v>-1$, and
$\operatorname{Re}\beta>0$, the flux vanishes: $S(0)=0$, the factor
$1-t_i$ kills the endpoint $t_i=1$, and on a collision face
$f_i-f_j=O(t_i-t_j)$ while the measure supplies
$|t_i-t_j|^{2\operatorname{Re}\beta}$.  On $a=a_0$, integrating
Eq.~\eqref{eq:S-totalder} proves Eq.~\eqref{eq:S-intertwining}; generic complex
parameters follow degree by degree by meromorphic continuation.

Finally insert Eq.~\eqref{eq:S-Psiexpand} into
Eq.~\eqref{eq:S-intertwining}.  Equations~\eqref{eq:S-eigen} and
\eqref{eq:S-Rmatrix} give
\begin{equation}
 \boxed{
 \kappa_YF_Y^{\rm Sel}={1\over2}\sum_{W=Y-x}
 (\phi_x+m_1)(\phi_x+m_2)M_{Y/W}F_W^{\rm Sel}.}
 \label{eq:S-recursion}
\end{equation}

\section{Solution of the recursion and the Selberg theorem}
\label{app:factorization}

Define
\begin{equation}
 \tau_\rho(z)=\beta^{-|\rho|}
 \prod_{s\in\rho}[z+\operatorname{ct}(s)]
 \label{eq:S-tau}
\end{equation}
and, without assuming it equals the integral, define
\begin{align}
 F_{\lambda,\mu}^{\rm cand}={}&(-1)^{|\lambda|+|\mu|}
 \tau_\lambda(-v-\beta n)
 \notag\\
 &\times\tau_\lambda(-u-v-\beta n-1+\beta)
 \notag\\
 &\times\tau_\mu(\beta n)
 \tau_\mu(u+\beta n+1-\beta).
 \label{eq:S-candtau}
\end{align}
The relation between $a$ and the Selberg parameters in
Eq.~\eqref{eq:S-Selbergpars}, together with
Eq.~\eqref{eq:S-masses}, implies
\begin{equation}
 F_Y^{\rm cand}=(-1)^{|Y|}\prod_{s\in Y}
 { (\phi_s+m_1)(\phi_s+m_2)\over\beta^2}.
 \label{eq:S-candbox}
\end{equation}
For $Y=W+x$, the old box factors cancel identically, leaving the
division-free relation
\begin{equation}
 (\phi_x+m_1)(\phi_x+m_2)F_W^{\rm cand}
 =-\beta^2F_Y^{\rm cand}.
 \label{eq:S-boxratio}
\end{equation}
Using Eq.~\eqref{eq:S-boxratio} in the right-hand side of
Eq.~\eqref{eq:S-recursion}, and then using the row sum
Eq.~\eqref{eq:S-rowsum}, gives
\begin{align}
 \mathrm{RHS}
 &=-{\beta^2\over2}F_Y^{\rm cand}
 \sum_{W=Y-x}M_{Y/W}
 =\kappa_YF_Y^{\rm cand}.
 \label{eq:S-candrec}
\end{align}
Hence the actual Selberg average and the candidate obey the same triangular
recursion and have the same vacuum value,
$F_\vac^{\rm Sel}=F_\vac^{\rm cand}=1$.  For generic parameters
$\kappa_Y\ne0$, Eq.~\eqref{eq:S-recursion} determines level $|Y|$ from level
$|Y|-1$ uniquely.  Equality at generic parameters therefore follows by
induction, and equality on special parameter loci follows by meromorphic
continuation.  We have proved
\begin{align}
 &\left\langle
 J_{\lambda,\mu}\left(a,-p_k-{v\over\beta},p_k\right)
 \right\rangle_S
 \notag\\
 &\quad=(-1)^{|\lambda|+|\mu|}
 \tau_\lambda(-v-\beta n)
 \tau_\lambda(-u-v-\beta n-1+\beta)
 \notag\\[-1mm]
 &\qquad\quad\times\tau_\mu(\beta n)
 \tau_\mu(u+\beta n+1-\beta).
 \label{eq:S-finalSelberg}
\end{align}
This is Eq.~(35) of Ref.~\cite{MorozovSmirnov}, now derived at all levels.

For completeness, the screening number in the preceding derivation initially
has $n\in\mathbb Z_{\geq0}$.  At any fixed total degree $d$, however, the
insertion in Eq.~\eqref{eq:S-finalSelberg} is a finite symmetric polynomial in
the power sums $p_k$.  Expanding it in the ordinary Jack basis
$P_\rho^{(1/\beta)}$, $|\rho|\leq d$, and using the normalized
Kadell--Selberg evaluation~\cite{Kadell}, expresses its average as a finite
sum of products of generalized Pochhammer symbols.  Consequently each
coefficient, after imposing $a=a_0(n)$, is a rational function of $n$ and a
meromorphic function of $u,v,\beta$.  Since Eq.~\eqref{eq:S-finalSelberg} holds for every admissible
nonnegative integer $n$ in the convergence domain, equality of these rational
functions defines its unique coefficientwise continuation to complex $n$.
The same finite-degree argument extends across the remaining parameter
hypersurfaces by meromorphic continuation.  This is the continuation used for
the physical values $n_\pm$ below; no noninteger-dimensional integral is
being introduced.

\section{DF--Nekrasov map and termwise AGT}
\label{app:agt-map}

The generalized Cauchy identity expands the interaction between the two DF
screening sectors as
\begin{align}
 &\exp\!\left[\beta\sum_{k\ge1}\frac{\Lambda^k}{k}
 \left\{p_k\left(-q_k-\frac{v_-}{\beta}\right)\right.\right.\notag\\[-1mm]
 &\hspace{34mm}\left.\left.+q_k\left(-p_k-\frac{v_+}{\beta}\right)
 \right\}\right]\notag\\
 &\quad=\sum_{\lambda,\mu}\Lambda^{|\lambda|+|\mu|}
 \frac{J_{\lambda,\mu}(a,-p_k-v_+/\beta,p_k)}{e(T_{\lambda,\mu})}
 \notag\\[-1mm]
 &\hspace{31mm}\times
 J^*_{\lambda,\mu}(a,q_k,-q_k-v_-/\beta).
 \label{eq:S-DFCauchy}
\end{align}
This is the generalized-Jack Cauchy expansion used in
Ref.~\cite{MorozovSmirnov}.  The denominator is the tangent-space Euler class
\cite{NekrasovOkounkov,SchiffmannVasserot,MorozovSmirnov}
\begin{equation}
 e(T_{\lambda,\mu})=
 e_{\lambda,\lambda}(0)e_{\lambda,\mu}(2a)
 e_{\mu,\lambda}(-2a)e_{\mu,\mu}(0),
 \label{eq:S-tangent}
\end{equation}
with $e_{\lambda,\mu}$ defined in Eq.~\eqref{eq:S-e-factor}.  Averaging the
two screening sectors independently gives
\begin{equation}
 \mathcal B_{\rm DF}(\Lambda)=\sum_Y\Lambda^{|Y|}
 {F_Y^+F_Y^{-,*}\over e(T_Y)}.
 \label{eq:S-blocksum}
\end{equation}
The generalized Cauchy identity naturally uses the
$\mathrm{Vir}\otimes\mathcal H$ (equivalently, $U(1)$-dressed) vertex
normalization.  Relative to the conventional pure Virasoro block
$\mathcal F_{\rm Vir}$,
\begin{equation}
 \mathcal B_{\rm DF}(\Lambda)=Z_{U(1)}(\Lambda)\mathcal F_{\rm Vir}(\Lambda),
 \qquad
 Z_{U(1)}(\Lambda)=(1-\Lambda)^\nu,\qquad
 \nu=\frac{v_+v_-}{2\beta}.
 \label{eq:S-U1-factor}
\end{equation}
This is the free-boson contraction between the insertions at $\Lambda$ and
$1$: with $v_+=2b\alpha_2$, $v_-=2b\alpha_3$, and $b^2=\beta$, its exponent
is $2\alpha_2\alpha_3$~\cite{AGT,AFLT}.
The first sector obeys
$a=-\beta n_+-(u_++v_++\varepsilon)/2$.  In the conventions of
Eqs.~\eqref{eq:S-dual} and \eqref{eq:S-Jstar}, $e_Y$ is the strict dual,
whereas the Euler-normalized paper dual $J_Y^*=e(T_Y)e_Y$ satisfies the
component-reversal identity
\begin{equation}
 J^*_{\lambda,\mu}(a;x,y)=J_{\mu,\lambda}(-a;y,x).
 \label{eq:S-dual-reversal}
\end{equation}
as in Ref.~\cite{MorozovSmirnov}.
Thus the second sector reduces to the same Selberg theorem with
$a=+\beta n_-+(u_-+v_-+\varepsilon)/2$.  Applying
Eq.~\eqref{eq:S-finalSelberg} after this reversal gives
\begin{align}
 F_{\lambda,\mu}^{-,*}={}&(-1)^{|\lambda|+|\mu|}
 \tau_\lambda(\beta n_-)
 \tau_\lambda(u_-+\beta n_-+1-\beta)
 \notag\\
 &\times\tau_\mu(-v_--\beta n_-)
 \tau_\mu(-u_--v_--\beta n_--1+\beta).
 \label{eq:S-dualaverage}
\end{align}

Let the four hypermultiplet masses be $\mathfrak m_1,\ldots,\mathfrak m_4$
and the physical Coulomb modulus be $\mathfrak a$.  With the physical Omega
parameters introduced in Eq.~\eqref{eq:S-physical-beta}, the standard DF/AGT
parameter map~\cite{AGT,AFLT,MorozovSmirnov} is
\begin{align}
 \beta&=-\frac{\epsilon_1}{\epsilon_2},\qquad
 a=\frac{\mathfrak a}{\epsilon_2},\notag\\
 n_+&=\frac{\mathfrak a-\mathfrak m_2}{\epsilon_1},\qquad
 n_-=\frac{-\mathfrak a-\mathfrak m_4}{\epsilon_1},\notag\\
 u_+&=\frac{\mathfrak m_1-\mathfrak m_2-\epsilon_1-\epsilon_2}{\epsilon_2},
 \notag\\
 u_-&=\frac{\mathfrak m_3-\mathfrak m_4-\epsilon_1-\epsilon_2}{\epsilon_2},
 \notag\\
 v_+&=\frac{-\mathfrak m_1-\mathfrak m_2}{\epsilon_2},\qquad
 v_-=\frac{-\mathfrak m_3-\mathfrak m_4}{\epsilon_2}.
 \label{eq:S-dictionary}
\end{align}
In particular, the exponent in Eq.~\eqref{eq:S-U1-factor} is
\begin{equation}
 \nu=-\frac{(\mathfrak m_1+\mathfrak m_2)
 (\mathfrak m_3+\mathfrak m_4)}{2\epsilon_1\epsilon_2}.
 \label{eq:S-U1-physical}
\end{equation}
Substitution gives the eight $\tau$ arguments explicitly:
\begin{align}
 -v_+-\beta n_+
 &=\frac{\mathfrak a+\mathfrak m_1}{\epsilon_2},\notag\\
 -u_+-v_+-\beta n_+-1+\beta
 &=\frac{\mathfrak a+\mathfrak m_2}{\epsilon_2},\notag\\
 \beta n_+&=\frac{-\mathfrak a+\mathfrak m_2}{\epsilon_2},\notag\\
 u_++\beta n_++1-\beta
 &=\frac{-\mathfrak a+\mathfrak m_1}{\epsilon_2}.
 \label{eq:S-plusargs}
\end{align}
\begin{align}
 \beta n_-&=\frac{\mathfrak a+\mathfrak m_4}{\epsilon_2},\notag\\
 u_-+\beta n_-+1-\beta
 &=\frac{\mathfrak a+\mathfrak m_3}{\epsilon_2},\notag\\
 -v_--\beta n_-&=\frac{-\mathfrak a+\mathfrak m_3}{\epsilon_2},\notag\\
 -u_--v_--\beta n_--1+\beta
 &=\frac{-\mathfrak a+\mathfrak m_4}{\epsilon_2}.
 \label{eq:S-minusargs}
\end{align}
Restoring dimensions, Eqs.~\eqref{eq:S-plusargs} and
\eqref{eq:S-minusargs} show that the two Selberg averages supply precisely the
four matter Euler classes.  The signs $(-1)^{|Y|}$ cancel between the two
averages, and their normalization powers of $\beta$ cancel those in
$e(T_Y)$.  Therefore each fixed point satisfies
\begin{equation}
 \boxed{
 {F_Y^+F_Y^{-,*}\over e(T_Y)}
 ={\displaystyle\prod_{f=1}^4\prod_{s\in Y}
 (\Phi_s+\mathfrak m_f)\over e_{\rm phys}(T_Y)}
 =Z_Y^{\rm Nek}.}
 \label{eq:S-termwise}
\end{equation}
Summing Eq.~\eqref{eq:S-termwise} over all double partitions gives the exact
normalization chain
\begin{equation}
 \mathcal B_{\rm DF}=Z_{U(1)}\mathcal F_{\rm Vir}
 =Z_{\rm inst}^{U(2)},\qquad
 \mathcal F_{\rm Vir}=\frac{Z_{\rm inst}^{U(2)}}{Z_{U(1)}}
 \equiv Z_{\rm inst}^{SU(2)}.
 \label{eq:S-final-AGT-chain}
\end{equation}
Thus the termwise equality is with the trace-free $U(2)$ fixed-point sum
($a_1=-a_2$), while the pure $SU(2)$/Virasoro statement is obtained only
after the explicit factor \eqref{eq:S-U1-factor} is removed.

\section{Verification of the kernel identity}
\label{app:operator-ledger}

We complete the conjugation calculation used in
Eqs.~\eqref{eq:S-structures} and \eqref{eq:S-totalder}.  Introduce the
combined index $I=(X,k)$, where $X=S,R$, and write
\begin{equation}
 K=\frac12\sum_{I,J}A_{IJ}\partial_I\partial_J
 +\sum_I G_I\partial_I .
 \label{eq:H-K-tensor}
\end{equation}
The nonzero second-order coefficients, read from
Eq.~\eqref{eq:S-K-SR}, are
\begin{equation}
 \begin{aligned}
 A_{S_rS_s}&=rsS_{r+s},&
 A_{R_rR_s}&=rsS_{r+s},\\
 A_{S_rR_s}&=A_{R_sS_r}=rsR_{r+s}.
 \end{aligned}
 \label{eq:H-second-order}
\end{equation}
The first-order coefficients are
\begin{align}
 G_{S_k}={}&
 \frac{\beta k}{4}\sum_{r+s=k}(S_rS_s+R_rR_s)
 +\varepsilon k^2S_k
 -\left(ak+\frac{\varepsilon}{2}k^2\right)R_k,
 \notag\\
 G_{R_k}={}&
 \frac{\beta k}{4}\sum_{r+s=k}(S_rR_s+R_rS_s)
 +\left(\frac{\varepsilon}{2}k^2-ak\right)S_k .
 \label{eq:H-first-order}
\end{align}

Conjugation by the kernel shifts the derivatives by
\begin{equation}
 d_{S_k}=-\frac{v}{2k},
 \qquad
 d_{R_k}=\frac{\beta p_k+v/2}{k}.
 \label{eq:H-shifts}
\end{equation}
Consequently,
\begin{equation}
 k_0:=(\mathcal K^{-1}K\mathcal K)1
 =\frac12\sum_{I,J}A_{IJ}d_Id_J+\sum_I G_Id_I .
 \label{eq:H-k0}
\end{equation}

Write
\begin{equation}
 B=T+\varepsilon S_1-
 \left(a+\frac{\varepsilon}{2}\right)R_1,
 \qquad
 T=\sum_{k\geq1}k
 \left(S_{k+1}\partial_{S_k}
 +R_{k+1}\partial_{R_k}\right).
 \label{eq:H-B}
\end{equation}
Its scalar conjugate is
\begin{align}
 b_0:={}&(\mathcal K^{-1}B\mathcal K)1
 \notag\\
 ={}&
 \sum_{k\geq1}\left[
 -\frac v2S_{k+1}
 +\left(\beta p_k+\frac v2\right)R_{k+1}\right]
 +\varepsilon S_1
 -\left(a+\frac{\varepsilon}{2}\right)R_1 .
 \label{eq:H-b0}
\end{align}
Let $b_I=\partial_Ib_0$.  Since $b_0$ is linear in the variables
$S_k,R_k$, its second derivatives vanish.  Using $C=[K,B]$, the common
product $b_0k_0$ cancels and gives
\begin{equation}
 c_0:=(\mathcal K^{-1}C\mathcal K)1
 =\frac12\sum_{I,J}A_{IJ}(d_Ib_J+d_Jb_I)
 +\sum_I G_Ib_I-Tk_0 .
 \label{eq:H-c0}
\end{equation}

It follows from Eq.~\eqref{eq:S-L-a} that
\begin{equation}
 (\mathcal K^{-1}\mathcal L_a\mathcal K)1
 =k_0-\frac12c_0+\frac{v+\varepsilon}{2}b_0
 +\frac{(u-v)(u+v+2\varepsilon)}8S_1 .
 \label{eq:H-conjugated-L}
\end{equation}
Substituting Eqs.~\eqref{eq:H-k0}--\eqref{eq:H-c0}, imposing
$a=a_0$, and using
$f_\ell=-\frac{\beta}{2}(S_\ell-S_{\ell-1})$ yields
\begin{align}
 (\mathcal K^{-1}\mathcal L_{a_0}\mathcal K)1
 ={}&
 \sum_{\ell\geq1}
 (\ell+u-\beta\ell)f_\ell p_{\ell-1}
 \notag\\
 &+\beta\sum_{\ell\geq1}f_\ell
 \sum_{r=0}^{\ell-1}p_rp_{\ell-1-r}
 +\frac{\beta v}{2}\sum_{k\geq1}S_kp_k
 \notag\\
 &+\beta\sum_{\ell,k\geq1}
 f_\ell R_kp_{k+\ell-1}
 =\mathcal T_f ,
 \label{eq:H-final}
\end{align}
where $p_0=n$.  This is precisely the total-derivative expression in
Eq.~\eqref{eq:S-Tf}, and therefore verifies
Eq.~\eqref{eq:S-totalder} on the charge-balance hyperplane.

\ifdefined\ARXIVMERGE
\else

\end{document}
\fi

\end{document}